\documentclass[12pt]{iopart}

\usepackage{iopams}
\usepackage[pdftex]{graphicx}
\usepackage{dcolumn}
\usepackage{bm}
\usepackage{hyperref}
\usepackage{color}
\usepackage{booktabs,tabulary}
\usepackage{soul}
\usepackage[export]{adjustbox}
\usepackage{caption}
\usepackage{subcaption}
\usepackage{epstopdf}

\usepackage{multirow}
\usepackage[export]{adjustbox}
\usepackage{array}
\usepackage{makecell}
\usepackage{enumitem}

\usepackage{amssymb} 
\usepackage{bm} 

\newcommand{\spD}[1]{\fn{\tilde{\chi}_{_V}}{#1}}

\newcommand{\tens}[1]{\mathinner{\boldsymbol{#1}}}
\newcommand{\ten}[2]{\mathinner{#1_{#2}}}
\newcommand{\cL}{\mathinner{c_{{_L}_q}}}
\newcommand{\cT}{\mathinner{c_{{_T}_q}}}
\newcommand{\kL}{\mathinner{k_{{_L}_q}}}
\newcommand{\kT}{\mathinner{k_{{_T}_q}}}

\newcommand{\cLref}{\mathinner{c_{{_L}_1}}}
\newcommand{\kLref}{\mathinner{k_{{_L}_1}}}
\newcommand{\cTref}{\mathinner{c_{{_T}_1}}}
\newcommand{\kTref}{\mathinner{k_{{_T}_1}}}

\newcommand{\rT}{r_T}
\newcommand{\rL}{r_L}

\newcommand{\Hankel}[2]{\fn{\mathcal{H}_{#1}^{(1)}}{#2}}

\newcommand{\uvect}[1]{\hat{\vect{#1}}}

\setstcolor{blue}

\setul{0}{0.3ex}

\newcommand{\R}{\mathbb{R}}
\renewcommand{\d}[1]{\mathinner{d#1}}
\newcommand{\fn}[2]{\mathinner{#1\mathopen{\left(#2\right)}}}
\newcommand{\vect}[1]{{\bf #1}}
\newcommand{\E}[1]{\left\langle#1\right\rangle}

\usepackage[colorinlistoftodos,shadow,textwidth=18mm]{todonotes}

\usepackage{stackengine}
\setstackgap{S}{1pt}
\newcommand{\fvdots}{\Shortstack{. . . .}}

\newenvironment{cond}
	{\left\{
    \begin{array}{l l}
    }
    { 
    \end{array} 
    \right.
    }

\newenvironment{matrix_det}[1]
	{\left\vert\begin{array}{#1} }{ \end{array}\right\vert}

\newcommand{\eqref}[1]{(\ref{#1})}

\newcommand{\dd}[1]{\mathinner{\mathrm{d}#1}}

\newcommand{\Imag}{\mathrm{Im}}
\newcommand{\Real}{\mathrm{Re}}

\newcommand{\abs}[1]{\left\vert #1 \right\vert}

\begin{document}

\title[Effective Elastic Wave Characteristics of Composite Media]{Effective Elastic Wave Characteristics of Composite Media}

\author{J. Kim$^1$ and S. Torquato$^{1,2,3,4}$}
\address{$^1$Department of Physics, Princeton University, Princeton, New Jersey 08544, USA}
\address{$^2$Department of Chemistry, Princeton University, Princeton, New Jersey 08544, USA}
\address{$^3$Princeton Institute for the Science and Technology of Materials, Princeton University, Princeton, New Jersey 08544, USA}
\address{$^4$Program in Applied and Computational Mathematics, Princeton University, Princeton, New Jersey 08544, USA}

\ead{torquato@princeton.edu}
\vspace{10pt}
\begin{indented}
\item[]\today
\end{indented}

\begin{abstract}
We derive exact expressions for effective elastodynamic properties of two-phase composites in the long-wavelength (quasistatic) regime via homogenized constitutive relations that are local in space.
This is accomplished by extending the ``strong-contrast" expansion formalism that was previously applied to the static problem. 
These strong-contrast expansions explicitly incorporate complete microstructural information of the composite via an infinite set of $n$-point correlation functions.
Utilizing the rapid-convergence properties of these series expansions (even for extreme contrast ratios), we extract accurate approximations that depend on the microstructure via the spectral density, which is easy to compute or measure for any composite.
We also investigate the predictive power of modifications of such approximation formulas postulated elsewhere [J. Kim and S. Torquato, Proc.
Nat. Acad. Sci. {\bf 117}, 8764 (2020)] to extend their applicability {\it beyond the quasistatic regime.}  
The accuracy of these nonlocal microstructure-dependent approximations is validated by comparison to full-waveform simulation results for certain models of dispersions. 
We apply our formulas to a variety of models of nonhyperuniform and hyperuniform disordered composites. 
We demonstrate that hyperuniform systems are less lossy than their nonhyperuniform counterparts in the quasistatic regime, and stealthy hyperuniform media can be perfectly transparent for a wide range of wavenumbers.
Finally, we discuss how to utilize our approximations for engineering composites with prescribed elastic wave characteristics. 
\end{abstract}
\submitto{\NJP}

\section{Introduction}

	The theoretical determination of the effective elastic wave characteristics of multiphase composite media is of great importance in geophysics \cite{Biot1956, Guy1974, mavko1994relation}, exploration seismology \cite{Sheriff1995, sahimi_self-affine_2005}, diagnostic sonography \cite{Sarvazyan2013}, crack diagnosis \cite{Sutin1995, hamzehpour_acoustic_2016}, architectural acoustics \cite{Watson_RoomAcoustics} and acoustic metamaterials \cite{Yuan_2019_energy-harvesting}, among many examples. 
	Such effective elastic properties generally depend on the phase properties, phase volume fractions $\phi_i$, frequency $\omega$ or wavenumber $k_I$ of the incident elastic waves, and an infinite set of correlation functions that characterizes the composite microstructure \cite{Frank1964, Keller1966, jordan_effective_2015}.
	There have been numerous theoretical/computational attempts to estimate the effective elastic wave characteristics \cite{Frank1964, Keller1966, jordan_effective_2015,   keller_stochastic_1964, jing_acoustic_1992,  kerr_scattering_1992, sheng_introduction_2006, rohfritsch_influence_2020}.
	However, the preponderance of previous closed-form approximation formulas for the effective elastodynamic properties apply only in the quasistatic regime \cite{jordan_effective_2015, kerr_scattering_1992}, i.e., applicable when $k_I \ell \ll 1$, where $\ell$ is a characteristic heterogeneity length scale\footnote{Some multiple-scattering approximations for effective elastic waves are accurate beyond the quasistatic regime; see Ref. \cite{rohfritsch_influence_2020} and references therein.
However, these formulas require complicated scattering coefficients of individual scatterers.}, and under restrictive conditions.  
	One such closed-form approximation is the Gaunaurd-\"{U}berall approximation \cite{gaunaurd_resonance_1983, kerr_scattering_1992}, which we employ to compare to simulation data and our nonlocal formulas described below.   

\begin{figure}[h]
\center
\includegraphics[width = 0.5\textwidth]{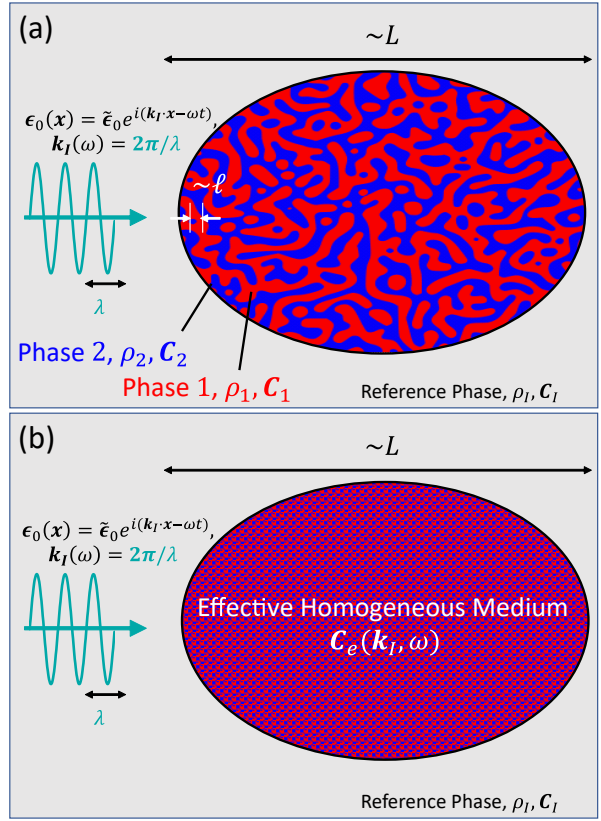}
\caption{
(a) Schematic of a large ellipsoidal, macroscopically anisotropic two-phase composite medium embedded in an infinite reference phase of mass density $\rho_I$ and stiffness tensor $\tens{C}_I$ (gray regions) under an applied elastic waves $\fn{\tens{\epsilon}_0}{\vect{x}} = \tilde{\tens{\epsilon}}_0 \fn{\exp}{i(\vect{k}_I \cdot\vect{x} - \omega t)}$ of frequency $\omega$.
The wavelength $\lambda$ associated with the applied wave can span from the quasistatic regime ($2\pi \ell /\lambda\ll 1$) down to the intermediate-wavelength regime ($2\pi \ell /\lambda \lesssim 1$), where $\ell$ is a characteristic length scale.
(b) After homogenization, the same ellipsoid can be regarded to be a specimen of a homogeneous medium with an effective stiffness tensor $\fn{\tens{C}_e}{\vect{k}_I,\omega }$, which depends on $\omega$ and $\vect{k}_I$. 
As noted in the main text, we omit the $\omega$ dependence of $\fn{\tens{C}_e}{\vect{k}_I,\omega }$ because (without loss of generally) we assume a linear dispersion relation between $\abs{\vect{k}_I}$ and $\omega$.
In the infinite-volume limit, we show that the effective wave characteristics are independent of the shape of the ellipsoidally-shaped composite. 
\label{fig:schematic}
}
\end{figure}

	Our focus in this paper is the theoretical determination of the effective dynamic stiffness tensor $\fn{\tens{C}_e}{\vect{k}_I,\omega}$ of a two-phase elastic composite in $d$-dimensional Euclidean space $\R^d$, which depends on the frequency $\omega$ or wavevector $\vect{k}_I$ of the incident elastic waves beyond the quasistatic regime; see \fref{fig:schematic}. 
	From this effective property, one can determine the corresponding effective wave speeds $c_e^{L,T}$ and attenuation coefficients $\gamma_e^{L,T}$.
	To achieve this goal, we first generalize the {\it strong-contrast} expansion formalism that has been employed to treat the static elastic problem \cite{torquato_exact_1997, Torquato1997, Torquato_RHM} to the elastodynamic problem in the quasistatic regime by establishing homogenized constitutive relations that are {\it local in space}.
	Because of the interplay between longitudinal and transverse waves and the complexity of the fourth-rank tensors that are involved, this task is considerably more challenging than the derivation of its electromagnetic counterparts \cite{Rechtsman2008, torquato_nonlocal_2020}.  
	The terms of the resulting quasistatic strong-contrast expansions are explicitly given in terms of integrals over products of Green's functions and the $n$-point correlation functions $\fn{S_n^{(i)}}{\vect{x}_1,\cdots,\vect{x}_n}$ of the random two-phase medium to infinite order. 
	Here, the quantity $\fn{S_n^{(i)}}{\vect{x}_1,\cdots,\vect{x}_n}$ gives the probability of finding $n$ points at positions $\vect{x}_1, \cdots, \vect{x}_n$ simultaneously in phase $i$. 	
	This implies that multiple scattering to all orders is exactly treated in the long-wavelength or quasistatic regime.
	It is noteworthy that the strong-contrast expansions are given in terms of expansion parameters that are rational functions of the phase moduli. 
	This endows strong-contrast expansions with rapid convergence properties, even for large phase contrast ratios. This behavior is to be distinguished from standard perturbation treatments that result in so-called ``weak-contrast" expansions \cite{Torquato_RHM} that slowly converge and only apply for small phase contrast ratios.

	Due to the fast-convergence properties of strong-contrast expansions, their lower-order truncations yield accurate closed-form approximate formulas for the effective dynamic moduli that apply to a wide class of microstructures.
	Postulated nonlocal variants of these formulas are resummed representations of the strong-contrast expansions that still accurately capture multiple scattering to all orders via the microstructural information embodied in the spectral density $\spD{\vect{Q}}$. 
 	The quantity $\spD{\vect{Q}}$ is the Fourier transform of the autocovariance
function $\fn{\chi_{_V}}{\vect{r}}\equiv \fn{S_2^{(i)}}{\vect{r}}-{\phi_i}^2$, where $\vect{r}\equiv \vect{x}_2-\vect{x}_1$, which can be easy to ascertain for general microstructures theoretically, computationally, or via scattering experiments \cite{Debye1949}.

	We also verify the accuracy of the postulated approximations via full-waveform simulations for certain benchmark models. 
	This validation allows us to use them to predict the effective elastic wave characteristics accurately well beyond the quasistatic regime for a wide class of
composite microstructures without computationally expensive full-blown simulations.
	As discussed in \sref{sec:local-str-cont}, such a broad microstructure class includes particulate composites consisting of identical or polydisperse particles of arbitrary shapes (ellipsoids, cylinders, polyhedra) that may or not overlap, cellular networks as well as systems without well-defined inclusions.
	Such broad applicability is a notable advantage of our formulas over other multiple-scattering approximations, such as Keller's approximation \cite{keller_stochastic_1964, derode_influence_2006}\footnote{Keller's approximation is derived for the simplified case in which only longitudinal waves propagate in a  very special system: colloidal suspensions of spherical particles in which the fluid has zero shear modulus.
	Such systems can be treated with the scalar Helmholtz equation, which is to be contrasted with our treatment of the full elastodynamic equations for macroscopically anisotropic media. }.
	Thus, our postulated formulas can be employed to accelerate the discovery of novel elastodynamic composites by appropriate tailoring of the spectral densities \cite{torquato_disordered_2016, Chen2017} and then generating the microstructures satisfying them \cite{Chen2017}, as elaborated in \sref{sec:discussion}. 

	While our strong-contrast formulas for the effective dynamic elastic moduli can be applied to periodic two-phase media, the primary applications are {\it spatially correlated} disordered media because they can provide advantages over periodic ones with high crystallographic symmetries \cite{Lopez2018,yu_bloch-like_2015}, including perfect isotropy and robustness against defects \cite{Florescu2013,man_isotropic_2013}.
We are interested in  both ``garden-variety" models \cite{Torquato_RHM} as well as exotic {\it hyperuniform} forms \cite{Torquato2003_hyper, Zachary2009a, Torquato2018_review} of disordered two-phase media. 
Hyperuniform two-phase systems are characterized by an anomalous suppression of volume-fraction fluctuations in the infinite-wavelength limit \cite{Torquato2003_hyper,Zachary2009a,Torquato2018_review}, i.e.,
the spectral density $\spD{\vect{Q}}$ obeys the condition 
\begin{equation}\label{eq:HU_condition}
\lim_{\abs{\vect{Q}}\to 0 }\spD{\vect{Q}} = 0.
\end{equation}
    Such hyperuniform two-phase media encompass all periodic systems, many quasi-periodic media, and exotic disordered ones; see Ref. \cite{Torquato2018_review} and references therein.
Disordered hyperuniform systems are exotic states of matters that lie between crystals and liquids; they behave like crystals in the way they suppress large-scale density fluctuations and yet are like liquids because they are statistically isotropic without any Bragg peaks  \cite{Torquato2003_hyper, Zachary2009a, Torquato2018_review}.
Hyperuniform systems have attracted considerable attention over the last decade because of their close connections to a broad spectrum of topics that arise in physical \cite{Lopez2018, Torquato2018_review, Torquato2015_stealthy, zhang_perfect_2016, Hexner2017_2, Ricouvier2017, Oguz2017, Ma2017,  yu_disordered_2018, wang_hyperuniformity_2018,  lei_hydrodynamics_2019, gorsky_engineered_2019, klatt_cloaking_2020, Jiao2014_chickenEyes, zheng_hyperuniformity_2020}, mathematical \cite{Ghosh2017_holeConjecture, brauchart_hyperuniform_2019, torquato_hidden_2019}, and materials sciences \cite{Chen2017, ma_3d_2016, xu_microstructure_2017, torquato_multifunctional_2018, kim_new_2019} as well as the emerging technological importance of the disordered varieties \cite{man_isotropic_2013, Torquato2018_review, gorsky_engineered_2019, Florescu2009,  ma_3d_2016, leseur_high-density_2016, Scheffold2017, klatt_characterization_2018, zhang_luneburg_2018}.

We apply our nonlocal strong-contrast formulas to predict the real and imaginary parts of the effective elastic moduli for model microstructures that possess some typical disorder (nonhyperuniform) as well as those with exotic hyperuniform disorder (\sref{sec:models}).
We are particularly interested in exploring the elastic properties of a special class of hyperuniform composites called disordered \textit{stealthy hyperuniform} media, which are defined to be those that possess zero-scattering intensity for a set of wavevectors around the origin \cite{Chen2017, Torquato2015_stealthy,Uche2004,Batten2008, Zhang2015}, i.e., 
\begin{equation}\label{eq:SHU-condition}
	\spD{\vect{Q}} = 0, ~\mathrm{for~} 0\leq Q\leq Q_\mathrm{U},
\end{equation}   
where $Q\equiv \abs{\vect{Q}}$.
Disordered stealthy hyperuniform materials have been shown to exhibit novel optical, acoustic, mechanical, and transport properties \cite{Rechtsman2008, xu_microstructure_2017, torquato_multifunctional_2018, leseur_high-density_2016, Zhang2016, deglinnocenti_hyperuniform_2016, gkantzounis_hyperuniform_2017,  kim_multifunctional_2020}.
Among other results, we show here that disordered hyperuniform media are generally less lossy than their nonhyperuniform counterparts. 
We also demonstrate that disordered stealthy hyperuniform particulate composites exhibit novel wave characteristics, including the capability to act as low-pass filters that transmit elastic waves isotropically without loss up to a selected wavenumber. 
Our results demonstrate that one can design the effective wave characteristics of a disordered composite, hyperuniform or not, by engineering spatial correlations of microstructure at prescribed length scales.

In \sref{sec:local-str-cont}, we present the strong-contrast formalism to derive corresponding expansions of the effective elastic wave characteristics of macroscopically anisotropic two-phase media in the quasistatic regime.
While we assume that both phases are elastically isotropic for simplicity, the effective elastic properties are described by a full fourth-rank tensor (what we mean by macroscopically anisotropic media) due to possibly statistically anisotropic microstructures. 
In \sref{sec:trun-approxs}, we extract strong-contrast approximations from the exact expansions.
In \sref{sec:modifications}, we extend the validity of the strong-contrast approximations for the effective dynamic moduli so that they apply well beyond the quasistatic regime. The accuracy of these nonlocal approximations is verified by comparison to full-waveform simulations for certain benchmark models.
In \sref{sec:models}, we describe four models of disordered composites that we treat in the paper, two of
which are nonhyperuniform and two
of which are hyperuniform.
In \sref{sec:predictions}, we investigate the microstructure-dependence of the effective elastic wave characteristics for these models.
Finally, we provide concluding remarks in \sref{sec:discussion}.

\section{Exact Strong-Contrast Expansions}
\label{sec:local-str-cont}

Here we extend the general strong-contrast formalism that was devised for the purely static elastic problem  \cite{torquato_exact_1997, Torquato1997, Torquato_RHM} to the elastodynamic problem in the long-wavelength (quasistatic) regime.
We first present a compact derivation of the expansions for the effective stiffness tensor $\fn{\tens{C}_e}{k_I,\omega}$ of a macroscopically anisotropic medium (\sref{sec:anisotropic}) and then specialize them to a macroscopically isotropic medium (\sref{sec:isotropic}).
Detailed derivations are given in the Supplementary Material (SM) \cite{supplementary_material}. 

We will exploit the same useful mathematical properties of the strong-contrast formalism that has been used to treat elastostatics \cite{Torquato1997, Torquato_RHM}, as we briefly outline here.
We begin with the integral solution of the elastodynamic equations in terms of the fourth-rank tensor Green's function.
The singular nature of the Green's function requires us to exclude a region around the singularity, but we recognize that the choice of the shape of this ``exclusion" region enables us to generate an infinite family of exact series expansions. 
By choosing a spherical exclusion region, we are able to derive strong-contrast expansions that rapidly converge, even for large phase contrast ratios.
The terms of the resulting strong-contrast expansions are explicitly given in terms of absolutely convergent integrals over products of Green's functions and certain $n$-point correlation functions through all orders. 
The rapid convergence of strong-contrast expansions enables us to extract accurate approximation formulas from the exact expansions. 

\subsection{Macroscopically Anisotropic Media} \label{sec:anisotropic}

Here, we consider macroscopically anisotropic two-phase media with isotropic phases but whose effective elastic properties are described by a full fourth-rank tensor $\tens{C}_e$. 
Macroscopic anisotropy arises with isotropic phases because the microstructure can generally be statistically anisotropic, e.g., layered media and oriented ellipsoids in a matrix. (See Ref. \cite{malyarenko_tensor-valued_2019} for a description of anisotropic phases for elastostatics.) 
We follow closely the strong-contrast formalism of Torquato \cite{Torquato1997, Torquato_RHM} but apply it to derive the analogous series expansions for the effective dynamic moduli.
We consider a macroscopically large ellipsoidal specimen of two-phase statistically homogeneous but anisotropic composite in $\R^d$ embedded inside an infinitely large reference phase $I$ with mass density $\rho_I$ and stiffness tensor $\tens{C}_I$; see \fref{fig:schematic}. 
The microstructure is perfectly general, and its inhomogeneity length scale $\ell$ is much smaller than the specimen size, i.e., $\ell \ll L$.
The shape of this specimen is purposely chosen as nonspherical since any rigorously correct expression for the effective property must ultimately be independent of the shape of the composite specimen in the infinite-volume limit.
Mathematically, we prove this below by showing that the strong-contrast formalism leads to effective properties that involve absolutely convergent integrals.

For a two-phase medium, we define the indicator function for phase $i$ as \cite{Torquato_RHM, Sahimi_HM1}  
\begin{equation}\label{eq:phase-indicator}
\fn{\mathcal{I}^{(i)}}{\vect{x}}
\equiv
\begin{cond}
1, & \vect{x} \mathrm{~lies~in~phase~}i\\
0, & \mathrm{ otherwise}
\end{cond},~\mathrm{for~}i=1,2.
\end{equation}
For statistically homogeneous media, its ensemble average is simply the phase volume fraction, i.e., $\phi_i \equiv \E{\fn{\mathcal{I}^{(i)}}{\vect{x}}}$ so that $\phi_1 + \phi_2 = 1$. 
The local stiffness tensor $\fn{\tens{C}}{\vect{x}}$ of such a medium can be written as
\begin{equation}
\fn{\tens{C}}{\vect{x}} \equiv
\tens{C}_1 \fn{\mathcal{I}^{(1)}}{\vect{x}} + \tens{C}_2 \fn{\mathcal{I}^{(2)}}{\vect{x}},
\label{eq:local_stiffness}
\end{equation}   
where $\tens{C}_i$ denotes the stiffness tensor of phase $i~(=1,2)$.
For simplicity, we take the reference phase to be phase $q$ (equal to 1 or 2).

In the ensuing discussion, we make the following three assumptions on the phase properties.
\begin{enumerate}[label=(\alph*)]
\item \label{assum1}
Phase 1 and phase 2 are elastically isotropic in $d$-dimensional Euclidean space $\R^d$, i.e.,  

\begin{equation}\label{eq:stiff_isotropic}
\tens{C}_i = d K_i \tens{\Lambda}_h + 2G_i \tens{\Lambda}_s,\quad(i=1,2)
\end{equation}
where $K_i$ and $G_i$ are bulk and shear moduli of phase $i~(=1,2)$, respectively.
Here, the {\it hydrostatic projection tensor} $\tens{\Lambda}_h$ and {\it shear projection tensor} $\tens{\Lambda}_s$ are constant fourth-rank tensors given by 
\begin{eqnarray}
\ten{(\Lambda_h)}{ijkl}& \equiv \frac{1}{d}\delta_{ij}\delta_{kl}, \label{def:hydrostatic}\\
\ten{(\Lambda_s)}{ijkl}& \equiv \frac{1}{2}\left( \ten{\delta}{ik}\ten{\delta}{jl}+\ten{\delta}{il}\ten{\delta}{jk}    \right) - \frac{1}{d}\delta_{ij}\delta_{kl},\label{def:shear}
\end{eqnarray}
where $\ten{\delta}{ij}$ is the Kronecker delta symbol.
The tensor $\tens{\Lambda}_h$ projects onto fields that are isotropic everywhere, whereas the tensor $\tens{\Lambda}_s$ projects onto fields that are trace-free (see the SM \cite{supplementary_material} for useful identities). 

\item \label{assum2}
Each phase is dissipationless, namely, the elastic moduli $K_i$ and $G_i$ for $i=1,2$ are real-valued and frequency-independent.

\item \label{assum3}
The mass densities of both phases are identical, i.e.,
\begin{equation}\label{eq:assum2}
\rho_1 = \rho_2 =\rho_e.
\end{equation}
\end{enumerate}
Assumption \ref{assum1} enables us to decompose the elastic waves in phase $i~(=1,2)$ into longitudinal and transverse waves with their respective wave speeds $c_{L_i}$ and $c_{T_i}$ \footnote{In this paper, `wave speed' always refers to the speed associated with the phase of the wave. This term is used instead of `phase speed' because `phase' in this paper refers to a constituent material of a composite.}, which are defined by
\begin{equation}
{c_{L_i}}^2 \equiv [K_i +2(1-1/d)G_i]/\rho_i, \quad {c_{T_i}}^2 \equiv G_i/\rho_i. \label{eq:speeds}
\end{equation}
Under this assumption, the Poisson ratio $\nu_i$ of phase $i$ is expressed as \cite{Torquato_RHM}
\begin{equation}\label{eq:Poisson ratio}
\nu_i = \frac{dK_i -2G_i}{d(d-1)K_i + 2G_i},
\end{equation}
and bounded in the range of $-1\leq \nu_i \leq 1/(d-1)$ \cite{Torquato_RHM, zohdi_introduction_2008}.
Assumption \ref{assum2} means that these speeds are independent of frequency $\omega$, implying the following linear dispersion relations: 
\begin{equation}\label{eq:disp-rel}
\omega = c_{L_i}/{k_{L_i}} = c_{T_i}/{k_{T_i}}\quad(i=1,2),
\end{equation}
where $k_{L_i}$ and $k_{T_i}$ are longitudinal and transverse wavenumbres, respectively. 
Assumption \ref{assum3} is achievable for many pairs of solid materials; see discussion in Ref. \cite{kim_multifunctional_2020}.

We suppose that the applied or incident elastic strain field $\fn{\tens{\epsilon}_0}{\vect{x}}$ is a plane wave of an angular frequency $\omega$, well-defined propagation direction $\uvect{k}$ in the reference phase, and the associated wavelength $\lambda$.
Our interest is in deriving an exact expression for the effective stiffness tensor $\fn{\tens{C}_e}{\omega}$ or, equivalently, $\fn{\tens{C}_e}{\kL}$ in the quasistatic regime ($\ell \ll \lambda$), where $\kL$ is the longitudinal wavenumber \eqref{eq:disp-rel} in the reference phase. 
While each phase is dissipationless, as stated in \ref{assum2}, the composite is generally lossy (i.e., $\tens{C}_e$ is complex-valued) due to scattering from the inhomogeneities.
Nonetheless, our results can be straightforwardly extended to viscoelastic media (with complex-valued moduli), but this will not be done in the present work.

Under the assumptions \ref{assum1}-\ref{assum3}, the local displacement field $\fn{\vect{u}}{\vect{x}}$ solves the time-harmonic wave equation:
\begin{equation}
\omega^2 u_i +  \left({\cL}^2-{\cT}^2   \right)\frac{\partial^2 u_k}{\partial x_i \partial x_k} + {\cT}^2 \frac{\partial^2 u_i}{\partial x_l \partial x_l} = -\frac{\partial \fn{\ten{P}{ij}}{\vect{x}}}{\partial x_j}, \label{eq:eq of motion isotropic}
\end{equation}
where the Einstein summation is implied, $\cL$ and $\cT$ are given in \eqref{eq:speeds}, and $\fn{\ten{P}{ij}}{\vect{x}}$ is the \textit{induced stress polarization field} given by
\begin{equation} \label{eq:P-tensor}
\fn{\tens{P}}{\vect{x}} \equiv {\rho_q}^{-1}\left[\fn{\tens{C}}{\vect{x}} - \tens{C}_q \right]:\fn{\tens{\epsilon}}{\vect{x}},
\end{equation}
and $\fn{\tens{\epsilon}}{\vect{x}}$ is the local strain tensor \cite{zohdi_introduction_2008}.
The symmetric second-rank tensor $\fn{\tens{P}}{\vect{x}}$ is the induced field relative to the reference phase $q$, and hence is nonzero only in the ``polarized" phase $p$ ($p\neq q$).

Following Torquato \cite{Torquato1997, Torquato_RHM}, we use a Green's function formalism to solve the wave equation \eqref{eq:eq of motion isotropic} for $\vect{u}$ for an arbitrary macroscopically anisotropic two-phase medium: 
\begin{eqnarray} \label{eq:int-eq-local-displacement}
\fn{\vect{u}}{\vect{x}}& = \fn{\vect{u}_0}{\vect{x}} + \int \fn{\tens{g}^{(q)}}{\vect{x}-\vect{x}'} \cdot \left [\nabla\cdot {\fn{\tens{P}}{\vect{x}'}} \right] \dd{\vect{x}'},
\end{eqnarray}  
where $\fn{\vect{u}_0}{\vect{x}}$ is related to the applied strain $\tens{\epsilon}_0$, and $\fn{\tens{g}^{(q)}}{\vect{r}}$ is the the second-rank Green's function associated with \eqref{eq:eq of motion isotropic}.  
Taking the symmetric part of the gradient of $\fn{\vect{u}}{\vect{x}}$ gives an integral equation for the local strain tensor $\fn{\tens{\epsilon}}{\vect{x}}$:
\begin{eqnarray}
\fn{\tens{\epsilon}}{\vect{x}} =& \fn{\tens{\epsilon_0}}{\vect{x}} + \int  \fn{\tens{G}^{(q)}}{\vect{x}-\vect{x}'}:\fn{\tens{P}}{\vect{x}'} \d{\vect{x}'} \label{eq:int-eq-strain},
\end{eqnarray}  
where the fourth-rank Green function $\fn{\tens{G}^{(q)}}{\vect{r}}$ associated with the reference phase $q$ is given by \cite{Torquato_RHM}
\begin{equation}\label{eq:four-rank-G}
\fn{\tens{G}^{(q)}}{\vect{r}} = -\tens{D}^{(q)} \fn{\delta}{\vect{r}} + \fn{\tens{H}^{(q)}}{\vect{r}},
\end{equation}
$\vect{r}\equiv\vect{x}-\vect{x}'$, $\tens{D}^{(q)}$ is a constant fourth-rank tensor that arises when one excludes an infinitesimal volume around the singularity of the Green function at $\vect{x}'=\vect{x}$, and $\fn{\tens{H}^{(q)}}{\vect{r}}$ is the contribution outside of this ``exclusion" region.

The fourth-rank tensor $\fn{\tens{H}^{(q)}}{\vect{r}}$ is symmetric under the following index changes, i.e.,
\begin{eqnarray}
\ten{H^{(q)}}{ijkl} = \ten{H^{(q)}}{jikl} = \ten{H^{(q)}}{ijlk} = \ten{H^{(q)}}{klij},
\end{eqnarray}
and its explicit expression is given by 
\begin{eqnarray}
\fl \fn{\tens{H}^{(q)}}{\vect{r}} =& \frac{-i\pi}{2 (2\pi)^{d/2}} \frac{1}{\omega^2 r^{d+2}} 
    \Big\{ 
    \left[\rL^{d/2+1}\Hankel{d/2+1}{\rL} - \rT^{d/2+1}\Hankel{d/2+1}{\rT}\right]\left(d\tens{\Lambda}_h + 2\tens{I}\right)
    + \rT^{d/2+2}\Hankel{d/2}{\rT} \tens{I}
    \nonumber \\
\fl
    &-\left[\rL^{d/2+2}\Hankel{d/2+2}{\rL}-\rT^{d/2+2}\Hankel{d/2+2}{\rT}\right] \left[2\fn{\tens{T}_1}{\vect{r}}+4\fn{\tens{T}_2}{\vect{r}} \right]
    -\rT^{d/2+3}\Hankel{d/2+1}{\rT}\fn{\tens{T}_2}{\vect{r}} 
    \nonumber \\
\fl &+\left[\rL^{d/2+3}\Hankel{d/2+3}{\rL} - \rT^{d/2+3}\Hankel{d/2+3}{\rT}\right]\fn{\tens{T}_3}{\vect{r}}  
\Big\}, \label{eq:H-tensor-explicit}
\end{eqnarray}
where $\rL \equiv \kL r$, $\rT \equiv \kT r$, $\tens{I}$ is the fourth-rank identity tensor, and $\Hankel{\nu}{x}$ is the Hankel function of the first kind of order $\nu$.
The three fourth-rank tensors $\fn{\tens{T}_i}{\vect{r}}$ for $i=1,2,3$ are defined, in component form, as 
\begin{eqnarray}
    \fn{\ten{(T_1)}{ijkl}}{\vect{r}} &\equiv \frac{1}{2} \left(\delta_{ij}\uvect{r}_k \uvect{r}_l + \uvect{r}_i \uvect{r}_j \delta_{kl}\right), \label{eq:T1}\\
    \fn{\ten{(T_2)}{ijkl}}{\vect{r}} &\equiv \frac{1}{4} \left(\uvect{r}_i\ten{\delta}{jk} \uvect{r}_l+\uvect{r}_j\ten{\delta}{ik} \uvect{r}_l + \uvect{r}_i\ten{\delta}{jl} \uvect{r}_k + \uvect{r}_j\ten{\delta}{il} \uvect{r}_k\right), \label{eq:T2}\\
    \fn{\ten{(T_3)}{ijkl}}{\vect{r}} &\equiv \uvect{r}_i \uvect{r}_j \uvect{r}_k \uvect{r}_l, \label{eq:T3}
\end{eqnarray}
$\uvect{r} \equiv \vect{r}/\abs{\vect{r}}$, and $\uvect{r}_i$ is the $i$th component of $\uvect{r}$.
Formulas for the traces of $\fn{\tens{H}^{(q)}}{\vect{r}}$ are provided in the SM \cite{supplementary_material}. 
    Note that \eqref{eq:H-tensor-explicit} reduces to its static counterpart given in Refs. \cite{torquato_exact_1997, Torquato1997, Torquato_RHM} up to a multiplicative factor $\rho_q$ in the static limit.   
    The constant tensor $\tens{D}^{(q)}$ depends on the ``exclusion-region" shape.
    Due to the fast-convergence properties of the resulting expansion discussed below and in Refs. \cite{torquato_exact_1997, Torquato_RHM, torquato_nonlocal_2020}, we choose a spherical-exclusion region, for which 
\begin{eqnarray}
\tens{D}^{(q)} &=  \frac{\rho_e\tens{\Lambda}_h}{dK_q + 2(d-1) G_q} + \frac{\rho_e d(K_q +2 G_q)\tens{\Lambda}_s}{G_q (d+2) [d K_q + 2(d-1)G_q]} \nonumber \\
& = \frac{1}{d {\cL}^2} \tens{\Lambda}_h    +\frac{1}{d+2}\left(\frac{2}{d {\cL}^2} +\frac{1}{{\cT}^2}\right) \tens{\Lambda}_s. \label{eq:D-tensor-sphr}
\end{eqnarray}
%

    The integral equation \eqref{eq:int-eq-strain} is written in a compact linear operator form as
    \begin{equation}\label{eq:strain-abstract}
    \tens{\epsilon} = \tens{\epsilon_0} + \tens{G}\tens{P}.
    \end{equation}
    Excluding the contribution from the exclusion region in \eqref{eq:strain-abstract}, we define \textit{generalized cavity strain} tensor: 
    \begin{equation} \label{eq:cavity-abstract}
    \tens{f} = \tens{\epsilon_0} + \tens{H}\tens{P}.
    \end{equation}
    Use of \eqref{eq:strain-abstract}, \eqref{eq:cavity-abstract}, and \eqref{eq:P-tensor} demonstrates that $\tens{P}$ and $\tens{f}$ are directly related as follows: 
    \begin{equation} \label{def:L-tensor_static}
    \fn{\tens{P}}{\vect{x}} = \left[\tens{L}^{(q)}\fn{\mathcal{I}^{(p)}}{\vect{x}}\right]:\fn{\tens{f}}{\vect{x}}, \quad(p\neq q)
    \end{equation}
    where 
\begin{eqnarray}
\tens{L}^{(q)} 
&\equiv
    \left(\tens{C}_p -\tens{C}_q\right)/\rho_q:\left[\tens{I}+\tens{D}^{(q)}:\left(\tens{C}_p-\tens{C}_q\right)/\rho_q\right]^{-1} \nonumber \\
&=
    d {\cL}^2 \left[    \kappa_{pq}\mathbf{\Lambda}_h + \frac{(d+2) {\cT}^2}{d{\cL}^2 + 2{\cT}^2} \mu_{pq} \mathbf{\Lambda}_s\right]\label{eq:L-tensor-sphr},
\end{eqnarray}
$\kappa_{pq}$ and $\mu_{pq}$ are the scalar polarizabilities for bulk and shear moduli, respectively, defined by
\begin{eqnarray} 
\kappa_{pq} & = \frac{K_p - K_q}{K_p + 2(d-1)G_q / d}, \label{eq:bulk_modulus_polarizability}\\
\mu_{pq} & = \frac{G_p - G_q}{ G_p + \left[dK_q/2 +(d+1)(d-2) G_q/d \right]G_q/(K_q +2 G_q)}.\label{eq:shear_modulus_polarizability}
\end{eqnarray}
Note that Eqs. \eqref{eq:D-tensor-sphr} and \eqref{eq:L-tensor-sphr} are identical to their static counterparts  \cite{Torquato1997, Torquato_RHM, torquato_effective_1998} up to a multiplicative factor $\rho_q$.

We now find a series expansion for the following homogenized constitutive relation
\begin{equation} \label{eq:homo-rel}
\fn{\E{\tens{P}}}{\vect{x}} = \fn{\tens{L}_e^{(q)}}{\kL}:\fn{\E{\tens{f}}}{\vect{x}},
\end{equation}
where $\E{\cdot}$ denotes an ensemble average, and the effective constant tensor $\fn{\tens{L}_e^{(q)}}{\kL}$ is explicitly given as
\begin{eqnarray}
\fn{\tens{L}_e^{(q)}}{\kL} 
= 
    \left[\fn{\tens{C}_e}{\kL} -\tens{C}_q\right]/\rho_e:\left\{\tens{I}+\tens{D}^{(q)}:\left[\fn{\tens{C}_e}{\kL}-\tens{C}_q\right]/\rho_e\right\}^{-1}. 
\end{eqnarray}
To do so, we solve $\fn{\tens{P}}{\vect{x}}$ in terms of $\tens{\epsilon}_0$ by iteratively substituting \eqref{eq:cavity-abstract} and \eqref{def:L-tensor_static}.
We then obtain the relation \eqref{eq:homo-rel} from the aforementioned expansion and an ensemble average of \eqref{eq:cavity-abstract} by eliminating $\tens{\epsilon_0}$ in order to avoid conditional convergence problems.

    Following the strong-contrast formalism of Torquato \cite{Torquato1997, Torquato_RHM}, we obtain an expression of the effective tensor $\tens{L}_e ^{(q)}$ in the form of series expansion:
    \begin{eqnarray}
\fl{\phi_p} ^2 \tens{L}^{(q)}:[\fn{\tens{L}_e ^{(q)}}{\kL}]^{-1} 
     &={\phi_p}^2\tens{L}^{(q)}: \left\{\tens{I}+\tens{D}^{(q)}:\left[\fn{\tens{C}_e}{\kL}-\tens{C}_q\right]/\rho_e \right\}:\left\{\left[\fn{\tens{C}_e}{\kL} -\tens{C}_q\right]/\rho_e \right\}^{-1}
    \nonumber\\
    &=\phi_p \tens{I} - \sum_{n=2}^\infty \fn{\tens{B}_n ^{(p)}}{\kL},\label{eq:exp-anisotropic}
\end{eqnarray}
where
\begin{eqnarray}
\fl\fn{\tens{B}_2 ^{(p)}}{\kL} =& \int_\epsilon \dd{\vect{x}_2} \fn{\tens{U}^{(q)}}{\vect{x}_1-\vect{x}_2} \fn{\chi_{_V}}{\vect{x}_1,\vect{x}_2}, \label{eq:B2} \\
\fl\fn{\tens{B}_n ^{(p)}}{\kL} =& (-1)^n (\phi_p)^{-(n-2)} \int_\epsilon \dd{\vect{x}_2} \cdots \dd{\vect{x}_n} \fn{\tens{U}^{(q)}}{\vect{x}_1-\vect{x}_2}:\fn{\tens{U}^{(q)}}{\vect{x}_2-\vect{x}_3}: \cdots: \fn{\tens{U}^{(q)}}{\vect{x}_{n-1},\vect{x}_n}\nonumber\\
&~~~~~~\times \fn{\Delta_n ^{(p)}}{\vect{x}_1,\vect{x}_2,\cdots,\vect{x}_n}, ~~n \geq 3, \label{eq:Bn}
\end{eqnarray}
$\fn{\tens{U}^{(q)}}{\vect{r}} \equiv \tens{L}^{(q)}:\fn{\tens{H}^{(q)}}{\vect{r}},$
and $\Delta_{n}^{(p)}$ is a position-dependent determinant involving up to the $n$-point correlation function associated with the dispersed phase $p$, i.e.,
\begin{equation}\label{eq:determinant}
\fl \fn{\Delta_{n}^{(p)}}{\vect{x}_1,\cdots,\vect{x}_n} =
\begin{matrix_det}{cccc}
\fn{S_2 ^{(p)}}{\vect{x}_1,\vect{x}_2}  &   \fn{S_1 ^{(p)}}{\vect{x}_1} & \cdots & 0 \\
\fn{S_3 ^{(p)}}{\vect{x}_1,\vect{x}_2,\vect{x}_3} & \fn{S_2 ^{(p)}}{\vect{x}_2,\vect{x}_3} & \cdots & 0 \\
\vdots & \vdots & \ddots & \vdots \\
\fn{S_n ^{(p)}}{\vect{x}_1,\cdots,\vect{x}_n} &\fn{S_{n-1} ^{(p)}}{\vect{x}_2,\cdots,\vect{x}_n} & \cdots & \fn{S_2 ^{(p)}}{\vect{x}_{n-1},\vect{x}_{n}} 
\end{matrix_det}.
\end{equation}
Here, $\fn{S_n^{(p)}}{\vect{x}_{1},\cdots, \vect{x}_{n}}$ is the $n$-point correlation function defined as
\begin{equation}\label{def:n-point correlation function}
\fn{S_n^{(p)}}{\vect{x}_1,\cdots, \vect{x}_n} \equiv \E{\fn{\mathcal{I}^{(p)}}{\vect{x}_1}\cdots \fn{\mathcal{I}^{(p)}}{\vect{x}_n}},
\end{equation}
which gives the probability for simultaneously finding $n$ points at $\vect{x}_1,\vect{x}_2,\cdots,\vect{x}_n$ in phase $p$ \cite{Torquato_RHM, Sahimi_HM1}.
Here, it is important to note that the integrals \eqref{eq:B2} and \eqref{eq:Bn} are absolutely convergent because while $\fn{\tens{H}^{(q)}}{\vect{r}}$ decays as $r^{-d}$ for large $r$, $\Delta_n^{(p)}$ identically vanishes at the boundary of the specimen \cite{Torquato_RHM}.
Therefore, this proves that the effective elastodynamic properties in the infinite-volume limit are independent of the shape of the ellipsoidal composite shown in Fig. \ref{fig:schematic}. 
The detailed derivation of the strong-contrast expansion \eqref{eq:exp-anisotropic} is given in the SM. 
Importantly, the exact series expansion \eqref{eq:exp-anisotropic} accounts for complete microstructural information (infinite set of $n$-point correlation functions) and hence multiple scattering to all orders in the quasistatic regime.

\vspace{10pt}
{\bf Remarks:}
\begin{enumerate}
\item Importantly, the strong-contrast expansion \eqref{eq:exp-anisotropic} is
a series representation of a linear fractional transformation
of the effective stiffness tensor $\fn{\tens{C}_e}{\kL}$ (left-hand side).
The series expansion in powers
of the polarizabilities $\kappa_{pq}$ and $\mu_{pq}$ of this particular rational function of $\fn{\tens{C}_e}{\kL}$ has important consequences for the predictive power of approximations derived from the expansion.
While this desirable feature is briefly discussed below, the reader is referred to Ref. \cite{torquato_nonlocal_2020} for detailed explanations for the corresponding electromagnetic problem.
We note that the strong-contrast formalism is a significant departure from standard perturbative expansions  that lead to ``weak-contrast" expansions in which the expansion parameters are simple differences in the phase moduli, implying that they converge only for small contrast ratios \cite{Torquato1997, Torquato_RHM}. 

%

\item The homogenized constitutive relation \eqref{eq:homo-rel} is {\it local in space} [i.e., $\fn{\E{\tens{P}}}{\vect{x}}$ at point $\vect{x}$ depends on   $\fn{\E{\tens{f}}}{\vect{x}}$ at the same position $\vect{x}$] and strictly valid in the long-wavelength regime.
In such a regime, the effective elastic moduli are independent of the direction of incident waves, as shown in the expansion \eqref{eq:exp-anisotropic}.
For shorter wavelengths, however, the associated relation must be nonlocal in space [i.e., $\fn{\E{\tens{P}}}{\vect{x}}$ at point $\vect{x}$ depends on   $\fn{\E{\tens{f}}}{\vect{x}'}$ at positions around $\vect{x}$], which can result in ``wavevector"-dependent effective elastic moduli, as was rigorously shown for the analogous electromagnetic
wave problem \cite{torquato_nonlocal_2020}.  

\item Note that the expansion \eqref{eq:exp-anisotropic} represents two different series: one for $q=1$ and $p=2$ and the other for $q=2$ and $p=1$.

\item In the static limit ($\omega\to 0$), the series \eqref{eq:exp-anisotropic} reduce to one derived for the static strong-contrast expansions \cite{torquato_exact_1997, Torquato1997, Torquato_RHM}.

\item In contrast to its static counterpart, the $n$-point microstructure-dependent tensors $\fn{\tens{B}_n^{(p)}}{\kL}$ given in Eqs. \eqref{eq:B2} and \eqref{eq:Bn} are functions of a frequency $\omega$ of the elastic waves or, equivalently, the longitudinal and transverse wavenumbers $\kL$ and $\kT$.
    Throughout this work, we use $\kL$ as an independent variable, instead of $\omega$ or $\kT$, for the following three reasons.
    First, the tensor $\fn{\tens{H}^{(q)}}{\vect{r}}$ as well as the effective stiffness tensors are conveniently written in terms of $\kL$.
    Second, $\kL$ is directly proportional to $\omega$ and $\kT$ [cf. \eqref{eq:speeds}].
    Furthermore, $\kL$ is directly related to a length scale, which is suitable for describing microstructural information rather than the temporal quantity $ \omega $.

\item The exact expansions \eqref{eq:exp-anisotropic} are independent of the reference phase $q$ and hence of the associated wavenumber $\kL$. 

\item Note that the strong-contrast formalism for the elastodynamic problem shares similar 
mathematical structure
to the electromagnetic counterpart \cite{Rechtsman2008, torquato_nonlocal_2020}. 
In both cases, the wave equations can be simplified to the Helmholtz equation [i.e., $\left(\nabla^2 + k^2 \right)\fn{\vect{u}}{\vect{x}} =0 $], which results in integral operator descriptions of their expansions being formally identical. 
However, there are important fundamental distinctions between the two problems.
Among other things, while electromagnetic waves have only transverse propagation modes, elastic waves always have both transverse and longitudinal modes with different wave speeds.
The interplay between these two propagation modes makes the theoretical determination of the effective elastodynamic properties generally more complex than its electromagnetic counterpart.

\end{enumerate}

\subsection{Macroscopically Isotropic Media} \label{sec:isotropic}

Here we assume that the composite is macroscopically isotropic.
In this case, the effective stiffness tensor $\tens{C}_e$ can be expressed in the effective bulk and shear moduli (denoted by $K_e$ and $G_e$, respectively).
Then, the series expansion \eqref{eq:exp-anisotropic} can be reduced to
\begin{equation}\label{eq:exp-isotropic}
{\phi_p}^2 \left[\frac{\kappa_{pq}}{\fn{\kappa_{eq}}{\kL}} \tens{\Lambda}_h + \frac{\mu_{pq}}{\fn{\mu_{eq}}{\kL}} \tens{\Lambda}_s\right] = \phi_p \tens{I} - \sum_{n=2}^\infty \fn{\tens{B}_n ^{(p)}}{\kL}.
\end{equation}
Utilizing the properties of two tensors $\tens{\Lambda}_h$ and $\tens{\Lambda}_s$ (see the SM \cite{supplementary_material}), one can separate \eqref{eq:exp-isotropic} into two expansions by taking the quadruple inner products of $\tens{\Lambda}_h$ and $\tens{\Lambda}_s$ with \eqref{eq:exp-isotropic}.
One is associated with the effective bulk modulus, and the other is related to the effective shear modulus:
\begin{eqnarray}
\fl\quad\fn{\kappa_{eq}}{\kL} &\equiv \frac{\fn{K_e}{\kL} -K_q}{\fn{K_e}{\kL} + 2(d-1)G_q /d} = \frac{{\phi_p}^2 \kappa_{pq}}{\phi - \sum_{n=2}^\infty \fn{C_n^{(p)}}{\kL}} ,& 
    \label{eq:exp-bulk}
\end{eqnarray}
\begin{eqnarray}
\quad \fl \fn{\mu_{eq}}{\kL} &\equiv \left[\fn{G_e}{\kL} -G_q\right]\Big\{\fn{G_e}{\kL} + \left[d K_q /2 +(d+1)(d-2)G_q /d\right]G_q /(K_q + 2G_q)\Big\}^{-1} 
    \nonumber \\
\fl &=\frac{{\phi_p}^2 \mu_{pq}}{\phi_p - \sum_{n=2}^\infty \fn{D_n^{(p)}}{\kL}} ,
    \label{eq:exp-shear}
\end{eqnarray}
respectively, where $\fn{C_n ^{(p)}}{\kL}\equiv \fn{\tens{B}_n ^{(p)}}{\kL}\fvdots\tens{\Lambda}_h$ and $\fn{D_n ^{(p)}}{\kL}\equiv 2[(d+2)(d-1)]^{-1}\fn{\tens{B}_n ^{(p)}}{\kL}\fvdots\tens{\Lambda}_s$.
Note that $\fn{C_n^{(p)}}{\kL}$ and $\fn{D_n^{(p)}}{\kL}$ involve the powers ${\kappa_{pq}}^m{\mu_{pq}}^{n-m}$, where an integer $m$ lies between $0$ and $n$.

Assuming that the composite is {\it passive} (i.e.,  it does not generate mechanical energy), and the time-harmonic factor of waves is $e^{-i\omega t}$, the imaginary parts of the effective elastic moduli must be non-positive, implying that
\begin{eqnarray*}
\Imag[\fn{K_e}{\kL}] \leq 0, ~\Imag[\fn{G_e}{\kL}] \leq 0,
\end{eqnarray*}
for any non-negative $\kL$.
In light of these properties, we have 
\begin{eqnarray}
\Imag\left[\kappa_{pq} \sum_{n=2}^\infty \fn{C_n^{(p)}}{\kL}\right] & \leq 0, \label{ineq:imag_Cn}
\\
\Imag\left[\mu_{pq}\sum_{n=2}^\infty \fn{D_n^{(p)}}{\kL}\right] &\leq 0. \label{ineq:imag_Dn}
\end{eqnarray} 

The effective elastic wave characteristics, including wave speeds $c_e^{L,T}$ and attenuation coefficients $\gamma_e^{L, T}$, are directly related to the effective moduli as follows: 
\begin{eqnarray}
c_e^L + i\gamma_e^L & \equiv \sqrt{[\fn{K_e}{\kL} + 2(1-1/d)\fn{G_e}{\kL}]/\rho_e}, \label{eq:eff-speed-L}\\
c_e^T+ i\gamma_e^T & \equiv \sqrt{\fn{G_e}{\kL}/\rho_e},\label{eq:eff-speed-T}
\end{eqnarray}
where $\rho_e = \rho_p = \rho_q$, and the superscripts $L$ and $T$ denote longitudinal and transverse waves, respectively.
Note that $\exp(-2\pi \gamma_e^L/c_e^L)$ and $\exp(-2\pi \gamma_e^T/c_e^T)$ represent the factors by which the amplitudes of the incident waves are attenuated inside the composite for a period of time $2\pi /\omega$.
    Thus, if $\gamma_e^L = \gamma_e^T = 0$ at certain wavenumbers (or frequencies), the composite is perfectly transparent, i.e., elastic waves propagate without any loss.  

\vspace{10pt}
{\bf Remarks:}
\begin{enumerate}
\item Any statistically isotropic medium is macroscopically isotropic, but the converse is not true.
For example, while cubic lattice packings are statistically anisotropic, they are macroscopically isotropic due to the cubic symmetry (see \sref{sec:vs-approxs}).

\item The dynamic strong-contrast expansions represented by \eqref{eq:exp-bulk} and \eqref{eq:exp-shear} possess fast-convergence properties for a wide class of microstructures, even at extreme phase contrast ratios (see Ref. \cite{torquato_nonlocal_2020} for detailed explanations).
Such convergence properties are attributed to the following two aspects.
First, even for extreme contrast ratios $K_p/K_q$ or $G_p/G_q$, the two expansion parameters $\kappa_{pq}$ and $\mu_{pq}$ are rational functions of the phase moduli and bounded by
\begin{eqnarray*}
-\infty < -\left[\frac{d^2(d-1)}{1+\nu_q}-d(d-1)^2\right]^{-1} \leq \kappa_{pq} < 1, \\
-\frac{2d}{(d-2)(d+1)} \leq \left[\frac{d}{2}-\frac{3(d+2)(2\nu_q -1)}{2d(5\nu_q -4)}\right]^{-1} \leq \mu_{pq} < 1,
\end{eqnarray*}
where $\nu_q$ is the Poisson ratio of the reference phase $q$.
Secondly, as Torquato \cite{torquato_exact_1997, Torquato1997} observed, the strong-contrast expansions in the static limit can be regarded to be ones that perturb around the wide class of optimal structures \cite{torquato_exact_1997, Torquato1997, torquato_effective_1998-1}, including the optimal multiscale Hashin-Shtrikman ``coated-spheres'' assemblages. The reader is referred to Refs. \cite{torquato_exact_1997, Torquato1997} for details.
It suffices to note here that such optimal two-phase media are characterized by a disconnected dispersed phase that is distributed throughout a connected matrix. 
These observations imply that the first few terms of the expansions \eqref{eq:exp-bulk} and \eqref{eq:exp-shear} can yield accurate approximations of the effective properties for a class of particulate composites as well as more general microstructures, even for extreme contrast ratios, provided that the dispersed phase is prevented from forming large clusters compared to the specimen size. 
Depending on whether the high-stiffness phase percolates or not, this broad microstructure class includes particulate media consisting of identical or polydisperse particles of general shape (ellipsoids, cubes, cylinders, polyhedra) that may or not overlap, cellular networks \cite{torquato_effective_1998-1} as well as media without well-defined inclusions. The reader is referred to Ref. \cite{torquato_nonlocal_2020} for a more detailed discussion of this issue.

\end{enumerate}

\section{Approximations at the Two-Point Level}\label{sec:trun-approxs}

Due to the fast-convergence properties of strong-contrast expansions, their truncations at low orders should yield accurate approximations for the effective bulk and shear moduli for the aforementioned wide class of microstructures over a broad range of volume fractions and contrast ratios; see also Ref. \cite{torquato_nonlocal_2020} for additional details. 
In what follows, we present such approximations by truncating the strong-contrast expansions after the second-order term.
The corresponding approximations at the three-point level are presented in \ref{app:3pt}. 
Detailed derivations are provided in Sec. I in the SM \cite{supplementary_material}.

Truncating \eqref{eq:exp-bulk} and \eqref{eq:exp-shear} at the two-point level and solving the left-hand sides of these truncated series for $K_e$ and $G_e$, respectively, yields 
\begin{eqnarray}
\quad\fl\frac{\fn{K_e}{\kL}}{K_q} =& 1+ \left[\frac{\cL^2}{\cL^2 - 2(1-1/d)\cT^2}\right]
\frac{{\phi_p}^2 \kappa_{pq}}{\phi_p (1-\phi_p \kappa_{pq}) - \fn{C_2^{(p)}}{\kL}}, \label{eq:Ke_two-point}
\end{eqnarray}
\begin{eqnarray}
\quad\fl\frac{\fn{G_e}{\kL}}{G_q} =& 1+ \left[\frac{d(d+2)\cL^2 /2}{d\cL^2 +2 \cT^2}\right]
\frac{{\phi_p}^2 \mu_{pq}}{\phi_p (1-\phi_p \mu_{pq})- \fn{D_2^{(p)}}{\kL}}\label{eq:Ge_two-point},
\end{eqnarray}
where $\fn{C_2^{(p)}}{\kL}$ and $\fn{D_2^{(p)}}{\kL}$ are defined respectively as
\begin{eqnarray}\label{eq:C2_explicit}
\fn{C_2^{(p)}}{\kL}& = \frac{\pi}{2^{d/2} \fn{\Gamma}{d/2}} \fn{\mathcal{F}}{\kL}\kappa_{pq} , \\
\label{eq:D2_explicit}
\fn{D_2^{(p)}}{\kL}& = \frac{\pi}{2^{d/2} \fn{\Gamma}{d/2}} \frac{d\cL^2 \fn{\mathcal{F}}{\kT} + 2\cT^2 \fn{\mathcal{F}}{\kL}}{d\cL^2 + 2\cT^2}\mu_{pq},
\end{eqnarray}
and the {\it local attenuation function} $\fn{\mathcal{F}}{Q}$ is defined as
\begin{eqnarray}
\fn{\mathcal{F}}{Q} & \equiv -\frac{2^{d/2}\fn{\Gamma}{d/2}}{\pi} {Q}^2 \int \frac{i}{4}\left(\frac{Q}{2\pi r}\right)^{\frac{d-2}{2}} \Hankel{d/2-1}{Q r} \fn{\chi_{_V}}{r}\dd{\vect{r}} \label{eq:local-F-real} \\
		& = -\frac{ \fn{\Gamma}{d/2}}{2^{d/2}\pi^{d+1}}{Q}^2 \int  \frac{\spD{\vect{q}}}{\abs{\vect{q}}^2 - Q^2}
\dd{\vect{q}}\label{eq:local-F-Fourier}, 
\end{eqnarray}
where $\fn{\Gamma}{x}$ is the gamma function, $\Hankel{\nu}{x}$ is the Hankel function of the first kind of order $\nu$, $\fn{\chi_{_V}}{r}\equiv \fn{S_2^{(p)}}{r}-{\phi_p}^2$ is the radial autocovariance function, and the {\it spectral density} $\spD{Q}$ is its Fourier transform.
Some important properties of $\fn{\mathcal{F}}{Q}$ are given in \ref{sec:prop-atten}.
Use of these properties of $\fn{\mathcal{F}}{Q}$ immediately shows that in the static limit ($\omega=0$), the parameters $\fn{C_2^{(p)}}{0}$ and $\fn{D_2^{(p)}}{0}$ are identically zero, which is consistent with previous studies \cite{torquato_effective_1998, Torquato_RHM}. 

Remarkably, $\fn{\mathcal{F}}{Q}$ also appears in the quasistatic strong-contrast approximations for the electromagnetic characteristics \cite{Rechtsman2008, torquato_nonlocal_2020}.
This commonality between the two wave problems at the two-point level allowed us to establish cross-property relations for the effective elastic and electromagnetic wave characteristics in Ref. \cite{kim_multifunctional_2020}.

\section{Improved Approximations at the Two-point Level}
\label{sec:modifications}

In order to extend the series expansions and approximations discussed in  \sref{sec:anisotropic}-\sref{sec:trun-approxs} beyond the quasistatic regime, one needs to generalize the strong-contrast expansion formalism to theories of elastodynamics that are nonlocal in space (see a recent review \cite{eltaher_review_2016}) from first principles, as we did for the electrodynamic problem in Ref. \cite{torquato_nonlocal_2020}.
Unlike the dielectric problems, however, such generalizations are nontrivial in the case of the elastodynamic problem because an elastically isotropic medium generally possesses multiple elastic wavenumbers at a given frequency $\omega$.

\subsection{Nonlocal Strong-Contrast Approximation}
 
Based on the following two observations, we postulate nonlocal strong-contrast approximations for the effective elastodynamic properties at the two-point level that are expected to be accurate beyond the quasistatic regime.
First, the local strong-contrast expansions for the elastodynamic and electromagnetic problems are similar in that the local attenuation function $\fn{\mathcal{F}}{Q}$, given by \eqref{eq:local-F-real}, appears in the local strong-contrast approximations of the effective dielectric constant that was rigorously derived in Ref. \cite{Rechtsman2008}; see also Ref. \cite{torquato_nonlocal_2020}. 
Second, guided by our exact formulation of the nonlocal effective electromagnetic characteristics \cite{torquato_nonlocal_2020}, such generalizations at the two-point level are tantamount to replacing the wavenumber-dependent local attenuation function $\fn{\mathcal{F}}{Q}$, defined in \eqref{eq:local-F-real}, with the wavevector-dependent {\it nonlocal attenuation function} $\fn{F}{\vect{Q}}$ defined by \cite{ torquato_nonlocal_2020, kim_multifunctional_2020}
\begin{eqnarray}
\fl\quad\fn{F}{\vect{Q}}&\equiv - \frac{2^{d/2} \fn{\Gamma}{d/2}}{\pi}{Q}^2 \int \frac{i}{4}\left(\frac{Q}{2\pi r}\right)^{d/2-1}
\Hankel{d/2-1}{Q r}  e^{-i\vect{Q} \cdot \vect{r}}\fn{\chi_{_V}}{\vect{r}} \dd{\vect{r}} \label{eq:nonlocal-F} 
\\
\fl &=
-\frac{ \fn{\Gamma}{d/2}}{2^{d/2}\pi^{d+1}}{Q}^2 \int  \frac{\spD{\vect{q}}}{\abs{\vect{q}+\vect{Q}}^2 - Q^2}
\dd{\vect{q}}.\label{eq:nonlocal-F-Fourier}
\end{eqnarray}
Unlike $\fn{\mathcal{F}}{Q}$, $\fn{F}{\vect{Q}}$ accounts for the contribution from spatial variation of the sinusoidal incident waves $\exp(-i\vect{Q}\cdot\vect{r})$ and thus more accurately estimates the scattering effects of waves associated with wavevector $\vect{Q}$ from the long- to intermediate-wavelength regimes.
(Important properties of $\fn{F}{\vect{Q}}$ for a statistically isotropic medium are provided in \ref{sec:prop-atten}.)
From these two observations, it is reasonable to assume that one can extend the range of applicable wavelengths by replacing $\fn{\mathcal{F}}{Q}$ in the local strong-contrast approximations at the two-point level [Eqs. \eqref{eq:Ke_two-point} and \eqref{eq:Ge_two-point}] with $\fn{F}{\vect{Q}}$, which are numerically verified in \sref{sec:vs-approxs}.
The resulting approximations are given respectively by 
\begin{eqnarray}
& \fl \frac{\fn{K_e}{\vect{\kL}}}{K_q}=& 1+ \left[\frac{\cL^2}{\cL^2 - 2(1-1/d)\cT^2}\right]
\frac{{\phi_p}^2 \kappa_{pq}}{\phi_p (1-\phi_p \kappa_{pq}) - \frac{\pi}{2^{d/2} \fn{\Gamma}{d/2}} \fn{F}{\vect{\kL}}\kappa_{pq}}, \label{eq:Ke_two-point-nonlocal}\\
& \fl \frac{\fn{G_e}{\vect{\kL}}}{G_q}=& 1+ \left[\frac{d(d+2)\cL^2 /2}{d\cL^2 +2 \cT^2}\right]
	\nonumber \\
&&\times  \frac{{\phi_p}^2 \mu_{pq}}{\phi_p (1-\phi_p \mu_{pq})- \frac{\pi}{2^{d/2} \fn{\Gamma}{d/2}} \frac{d\cL^2 \fn{{F}}{\vect{\kT}} + 2\cT^2 \fn{{F}}{\vect{\kL}}}{d\cL^2 + 2\cT^2}\mu_{pq}},	\label{eq:Ge_two-point-nonlocal}
\end{eqnarray}
where $\vect{\kL}$ and $\vect{\kT}$ are the longitudinal and transverse wavevectors of the incident waves, respectively, and $\fn{F}{\vect{Q}}$ is given in \eqref{eq:nonlocal-F}.
We emphasize that the nonlocal strong-contrast approximations for both elastic and electromagnetic properties share a common microstructure-dependent parameter $\fn{F}{\vect{Q}}$, which enabled us to establish cross-property relations linking those properties in Ref. \cite{kim_multifunctional_2020}. 
Note that, as we shown in a recent paper \cite{torquato_nonlocal_2020}, the analytic properties of $\fn{F}{\abs{\vect{Q}}}$ lead the nonlocal approximations \eqref{eq:Ke_two-point-nonlocal} and \eqref{eq:Ge_two-point-nonlocal} satisfies Kramers-Kronig relations for elastic waves \cite{booij_generalization_1982, ouis_frequency_2002}.
(These nonlocal approximations were first postulated in Ref. \cite{kim_multifunctional_2020} on physical grounds for establishing the cross-property relations.) 

For a statistically isotropic composite, as shown in \eqref{eq:Ke_two-point-nonlocal} and \eqref{eq:Ge_two-point-nonlocal}, the imaginary part of $\fn{F}{\abs{\vect{Q}}}$ directly determines the degree of attenuation, i.e., $\Imag[K_e]$ and $\Imag[G_e]$ or, equivalently, $\gamma_e^L$ and $\gamma_e^T$ defined in \eqref{eq:eff-speed-L} and \eqref{eq:eff-speed-T}.
In the quasistatic regime, assuming that the spectral density has the power-law scaling $\spD{Q}\sim Q^\alpha$, 
the effective attenuation coefficients $\fn{\gamma_e^{L,T}}{\kL}$ exhibit
\begin{eqnarray}\label{eq:atten-characteristics}
\fn{\gamma_e^{L,T}}{\kL} &\sim& 
\Imag[\fn{F}{\kL}] 
 \\
&\sim&
\begin{cond}
{\kL}^3, &\mathrm{nonhyperuniform~}(\alpha=0) \\
{\kL}^{3+\alpha}, &\mathrm{hyperuniform~}(\alpha>0)
\end{cond},
~\mathrm{as~}\kL \to 0^+, \nonumber
\end{eqnarray}
where nonhyperuniform systems take $\alpha=0$, whereas hyperuniform ones take $\alpha>0$ (see \ref{sec:prop-atten}).
Thus, hyperuniform media are less lossy than their nonhyperuniform counterparts as the wavenumber tends to zero.
Remarkably, the stealthy hyperuniform media are perfectly transparent up to a finite wavenumber:
\begin{equation} \label{eq:SHU-transparency}
\fn{\gamma_e^{L,T}}{\kL} = 0,\quad\mathrm{if~} 0\leq \kL \leq \frac{\cT}{\cL} \frac{Q_\mathrm{U}}{2},
\end{equation}
where $\cT/\cL = \sqrt{(1-2\nu_q)/[2(1-\nu_q)]}$, and $\nu_q$ is the Poisson ratio of the reference phase $q$.
 
\begin{figure*}
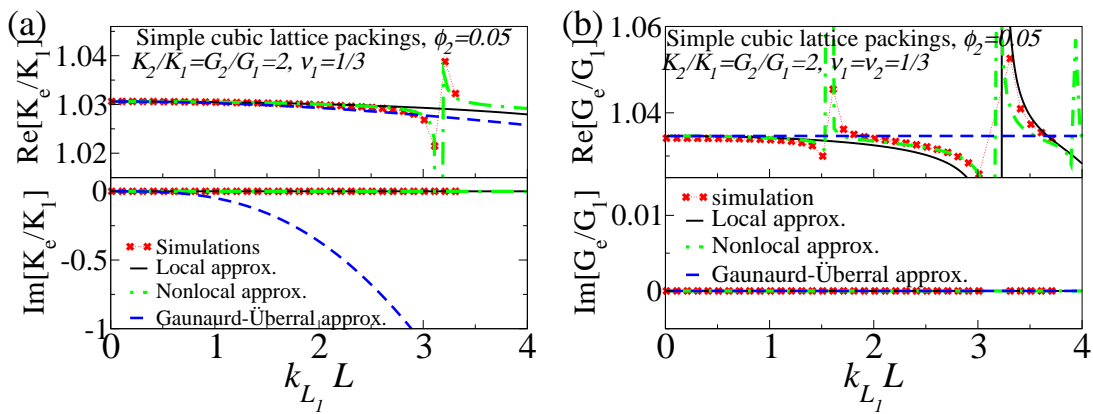

\subfloat{\includegraphics[width=0.45\textwidth]{Fig2a.eps}}
\hspace{5pt}
\subfloat{\includegraphics[width=0.45\textwidth]{Fig2b.eps}}
\caption{Comparison of the predictions of the local strong-contrast approximations [Eqs. \eqref{eq:Ke_two-point} and \eqref{eq:Ge_two-point}], the nonlocal variants [Eqs. \eqref{eq:Ke_two-point-nonlocal} and \eqref{eq:Ge_two-point-nonlocal}] and GUA [Eqs. \eqref{eq:GU_model_Ke} and \eqref{eq:GU_model_Ge}] for the effective dynamic bulk $\fn{K_e}{k_{_{L_1}}}$ and shear $\fn{G_e}{k_{_{L_1}}}$ moduli as functions of dimensionless waveunmber $ k_{_{L_1}} L$ of periodic packings to the corresponding simulation results.
We consider 3D cubic lattice packing of packing fraction $\phi_2=0.05$, contrast ratios $K_2/K_1 = G_2/G_1 = 2$, and Poisson ratio $\nu_1=1/3$.
Here, $k_{_{L_1}}$ is the longitudinal wavenumber in the reference phase (phase 1) along the $\Gamma$-$X$ direction, and $L$ is the nearest-neighbor distance.
\label{fig:simulations}
}
\end{figure*}

\subsection{Comparison of Simulations to Various Approximations}
\label{sec:vs-approxs}


Here we compare various approximations formulas for the effective dynamic elastic moduli to computer simulations, which are highly nontrivial calculations. 
In particular, we utilize our
fast-Fourier-transform (FFT) numerical
scheme presented elsewhere \cite{kim_multifunctional_2020}.
This procedure extends the one first devised for the effective static elastic moduli \cite{moulinec_numerical_1998} in order to treat elastodynamics.
The reader is referred to the SM \cite{supplementary_material} and Ref. \cite{kim_multifunctional_2020} for details.

In order to ensure convergence of the simulation procedure, we choose to study simple cubic lattice packings in which identical spheres of radius $a$ of phase 2 are embedded in the matrix phase (phase 1).
While the periodic packings are macroscopically isotropic, due to cubic symmetry, they are statistically anisotropic, implying that effective properties can depend on the direction of the incident wave $\vect{\kLref}$.
For simplicity, we only consider the case where $\vect{\kLref}$ is aligned with one of the minimal lattice vectors, i.e., $\Gamma$-$X$ direction in the first Brillouin zone.
Simple cubic lattice packings also provide stringent tests of the predictive power of the approximations at finite wavenumbers because they exhibit two salient and nontrivial elastic properties due to spatial correlations at intermediate length scales: transparency up to finite wavenumbers associated with the edges of
the first Brillouin zone (i.e., $\Imag[K_e]=0$ for $0\leq \kLref \lesssim \pi$ and $\Imag[G_e]=0$ for $0 \leq \kTref \lesssim \pi$), and resonance-like attenuation due to Bragg diffraction within the phononic bandgap (i.e., a peak in the imaginary parts or, equivalently, a sharp transition in the real parts).

We perform simulations for the case of simple cubic lattice of spheres in a matrix in which the packing fraction is $\phi_2=0.05$, contrast ratios are $K_2/K_1 = G_2/G_1 =2 $, and the Poisson ratio of the reference phase is $\nu_1=1/3$. 
In \fref{fig:simulations}, we compare the simulation results to the predictions from the strong-contrast approximations [Eqs. \eqref{eq:Ke_two-point} and \eqref{eq:Ge_two-point} for local approximations, and Eqs. \eqref{eq:Ke_two-point-nonlocal} and \eqref{eq:Ge_two-point-nonlocal} for the nonlocal counterparts] as well as the Gaunaurd-\"{U}berall approximation (GUA) [\eqref{eq:GU_model_Ke} and \eqref{eq:GU_model_Ge}].
While all approximations agree with the simulations in the quasistatic regime, the GUA and local strong-contrast approximations fail to capture properly two key features: no loss of energy up to finite wavenumbers and resonance-like attenuation in the band gaps.
However, the nonlocal strong-contrast approximations capture these two features and agree well with the simulation results, even beyond the quasistatic regime.
\vspace{10pt}
\section{Disordered Model Microstructures}
\label{sec:models}

\begin{figure*}
\begin{center}
\includegraphics[width = 0.9\textwidth]{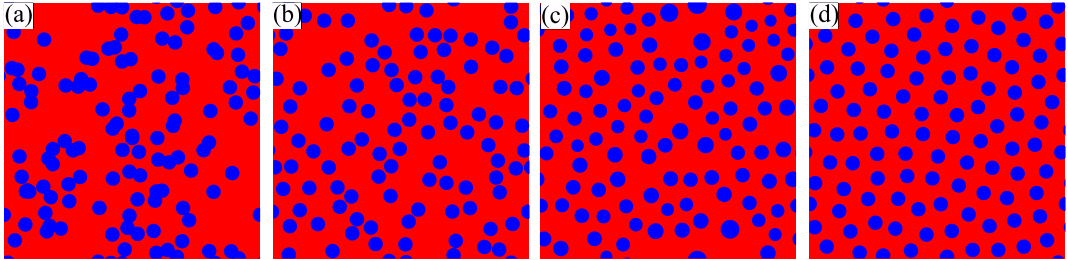}
\end{center}
\caption{Images of representative configurations of the four models of 2D disordered particulate media described in \sref{sec:models}.
While we focus on 3D models in this work, we present 2D images for the purpose of visualization. 
These include (a) overlapping spheres, (b) equilibrium (hard-sphere) packings, (c) class I hyperuniform polydisperse packings, and (d) stealthy hyperuniform packings.
The dispersed and matrix phases are shown in blue and red, respectively.
All models have an identical volume fraction of the disperse phase $\phi_2=0.25$.
Note that (a) and (b) are nonhyperuniform.
\label{fig:rep-images}
}
\end{figure*}

\begin{figure}[h]
	\begin{center}
	\includegraphics[width = 0.6\textwidth]{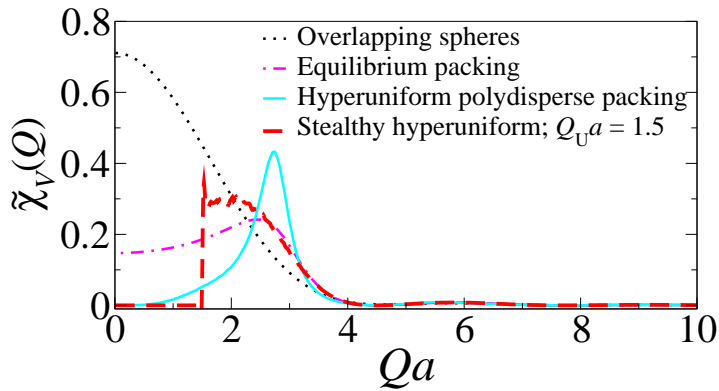}
	\end{center}
	\caption{(Color online) The spectral density $\spD{Q}$ as a function of dimensionless wavenumber $Qa$ for the four models of 3D disordered media at $\phi_2=0.25$: overlapping spheres, equilibrium packings, class I hyperuniform polydisperse packings, and stealthy hyperuniform packings.
For hyperuniform polydisperse packings, $a$ is the mean sphere radius.
The remaining models consist of identical spheres of radius $a$.
	\label{fig:spectral density}}
  \end{figure}
  
Here, we describe the four models of 3D disordered two-phase media that are statistically isotropic to study the microstructure-dependence of effective elastic properties.
The models include two nonhyperuniform systems (overlapping spheres and equilibrium packings) and two hyperuniform systems (class I hyperuniform polydisperse packings and stealthy hyperuniform packings).
In each mode, spherical particles of phase 2 are distributed throughout a matrix phase (phase 1).

\Fref{fig:rep-images} depicts representative images of these four models in two dimensions at the selected volume fraction of the disperse phase $\phi_2=0.25$ for the purpose of visualization. 
Note that the degree of volume-fraction fluctuations decreases from \fref{fig:rep-images}(a) to (d).
We also compute the corresponding spectral densities for these models in three dimensions and plot them in \fref{fig:spectral density}. 

	
	\subsection{Overlapping Spheres}
	\label{sec:overlapping}	
	Overlapping spheres (also called fully-penetrable-sphere model) refer to systems of identical spheres of radius $a$ whose centers are spatially uncorrelated in a matrix phase \cite{Torquato_RHM}.
	At a given number density $\rho$ in $d$-dimensional Euclidean space $\R^d$, the autocovariance function of this model can be written analytically \cite{ Torquato_RHM}.
	For $d=3$, it explicitly writes 
\begin{eqnarray}
\fl\fn{\chi_{_V}}{r} &= 
\exp\Bigg( -\rho  \fn{v_1}{a} \Big[2 \fn{\Theta}{x-1} 
+\left(1+\frac{3x}{2}-\frac{x^3}{2}\right) \fn{\Theta}{1-x}\Big]\Bigg)-{\phi_1}^2, \label{eq:autoco_OS}
\end{eqnarray}
where $\phi_1 = \exp(-\rho\fn{v_1}{a} )$ is the volume fraction of the matrix phase (phase 1), $v_1(a) =4\pi a^3/3$ is the volume of a sphere of radius $a$, $x\equiv r/(2a)$, and $\Theta(x)$ (equal to 1 for $x>0$ and zero otherwise) is Heaviside step function.
For $d=3$, the particle phase (phase 2) percolates when $\phi_2 \gtrsim 0.29$ (Ref. \cite{Rintoul1997}).
In this work, we consider this model for $\phi_2$ well below the percolation threshold.

	\subsection{Equilibrium Packings}\label{sec:equ-packing}
	\label{sec:HSF}
	Equilibrium packings are systems of identical hard spheres of radius $a$ in the (Gibbs) equilibrium distributions \cite{Torquato_RHM, Hansen_statmechReference}.
	Below freezing points, the Percus-Yevick solution \cite{Torquato_RHM, Hansen_statmechReference}, which is analytically solvable for odd values of $d$, well approximates the structure factors $\fn{S}{Q}$ of this model.
	For $d=3$, the analytic approximation of $\fn{S}{Q}$ is given by \cite{Torquato_RHM} 
	\begin{eqnarray}
\fn{S}{Q} = 
&\Big(1-\rho \frac{16 \pi  a^3 }{q^6} 
\Big\{\big[24 a_1 \phi_2 - 12 (a_1 + 2 a_2) \phi_2 q^2 
	\nonumber \\
	&\quad+ (12 a_2 \phi_2 + 2 a_1 + a_2\phi_2) q^4] \cos(q) 
	\nonumber \\
&+ [24 a_1 \phi_2 q - 2 (a_1 + 2 a_1 \phi_2 + 12 a_2 \phi_2) q^3\big] \sin(q)
	\nonumber \\
& -24 \phi_2 (a_1 - a_2 q^2) \Big\}\Big)^{-1},
\end{eqnarray}
where $q = 2Qa$, $a_1 = (1+2\phi_2)^2/(1-\phi_2)^4$, and $a_2 = -(1+0.5\phi_2)^2 /(1-\phi_2)^4$.
Using this solution in conjunction with the following formula \cite{Torquato_RHM, torquato_disordered_2016}
\begin{eqnarray}
  \spD{Q}= \rho \left( \frac{2\pi a}{Q} \right)^3 \fn{J_{3/2}^2}{Qa} \fn{S}{Q}
  \label{chi-packing}
\end{eqnarray}
yields the corresponding spectral density $\spD{Q}$.

\subsection{Hyperuniform Polydisperse Packings} \label{sec:tessell}

Class I hyperuniform sphere packings with a polydispersity in size can be constructed from nonhyperuniform progenitor point patterns via a tessellation-based procedure \cite{kim_new_2019, kim_methodology_2019}.
Specifically, we employ the centers of 3D configurations of equilibrium packings (\sref{sec:equ-packing}) at a packing fraction $0.45$ and particle number $N=1000$ as the progenitor point patterns.
We begin with the Voronoi tessellation \cite{Torquato_RHM} of these progenitor point patterns.
We then rescale the particle in the $j$th Voronoi cell $C_j$ without changing its center such that the packing fraction inside this cell is identical to a prescribed value $\phi_2<1$.
The same process is repeated over all cells.
The resulting packing fraction is $\phi_2 = \sum_{j=1}^N \fn{v_1}{a_j}/V_\mathfrak{F} = \rho \fn{v_1}{a}$, where $\rho$ is the number density of particle centers, $V_\mathfrak{F}$ is the volume of the periodic fundamental cell, and $a$ represents the mean sphere radius.
In the small-$\abs{\vect{Q}}$ regime, the spectral density of the resulting particulate composites exhibit a power-law scaling $\spD{\vect{Q}} \sim \abs{\vect{Q}}^4$,
which are of class I.
\subsection{Stealthy Hyperuniform Packings} \label{sec:stealthy}

	Stealthy hyperuniform packings of identical spheres, which are also class I, are defined by the spectral density vanishing around the origin; see  \eqref{eq:SHU-condition}.  
	We obtain the spectral density from their realizations for $d=3$ that are numerically generated via the following two steps.
Specifically, we first generate stealthy hyperuniform point configurations that include $N$ particles in a fundamental cell $\mathfrak{F}$ under periodic boundary conditions via the collective-coordinate optimization technique \cite{Uche2004, Batten2008, Zhang2015}, which amounts to finding numerically the ground-state configurations for the following potential energy;
\begin{equation}\label{eq:CC_potential}
\fn{\Phi}{\vect{r}^N} =\frac{1}{V_\mathfrak{F}} \sum_{\vect{Q}} \fn{\tilde{v}}{\vect{Q}}\fn{S}{\vect{Q}} +  \sum_{i <j} \fn{u}{r_{ij}},
\end{equation}
where $\fn{S}{\vect{Q}}$ is the structure factor of the particle centers,  
\begin{equation}
\fn{\tilde{v}}{\vect{Q}}
=
\begin{cond}
1, & 0< \abs{\vect{Q}} \leq Q_\mathrm{U} \\
0, &\mathrm{otherwise}
\end{cond}, \label{eq:cc-}
\end{equation}
and a soft-core repulsive term \cite{Zhang2017} 
\begin{equation} \label{eq:soft}
\fn{u}{r}
=
\begin{cond}
(1-r/\sigma)^2, & r < \sigma\\
0,&\mathrm{otherwise}
\end{cond}.
\end{equation}
Different from the usual procedure \cite{Uche2004, Batten2008, Zhang2015}, the interaction \eqref{eq:CC_potential} used here also includes a soft-core repulsive energy \eqref{eq:soft}, as was done in Ref. \cite{Zhang2017}.
	Thus, the resulting configurations are still disordered and stealthy hyperuniform, and their nearest-neighbor distances are larger than the length scale $\sigma$ due to the soft-core repulsion $\fn{u}{r}$.
	Finally, to create packings, we circumscribe the points by identical spheres of radius $a<\sigma/2$ under the constraint that they cannot overlap.
The parameters used to generate these packings are summarized in the SM \cite{supplementary_material}.

\subsection{Spectral Densities for the Four Models}\label{sec:chiv-models}

Here, we plot the spectral density $\spD{Q}$ for the fouar models for $d=3$ at a selected particle-phase volume fraction $\phi_2=0.25$.
From the long- to intermediate-wavelength regimes ($Qa \lesssim 4$), their spectral densities exhibit notable microstructure-dependence.
For example, overlapping spheres have the largest degree of volume-fraction fluctuations, followed by equilibrium packings.
By contrast, in the small-wavelength regime ($Qa \gg 4$), all four curves collapse onto a single curve because these models consist of spheres of similar sizes and thus have similar local structures.

\section{Predictions from Strong-Contrast Approximations}
\label{sec:predictions}

\begin{figure}
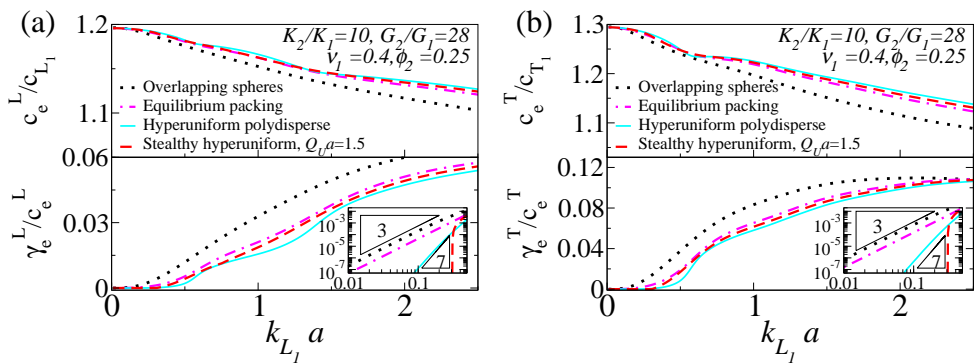

\begin{center}
\subfloat{\includegraphics[width=0.4\textwidth]{Fig5a.eps}}
\hspace{5pt}
\subfloat{\includegraphics[width=0.4\textwidth]{Fig5b.eps}}
\end{center}
\caption{
Predictions of the approximations \eqref{eq:Ke_two-point-nonlocal} and \eqref{eq:Ge_two-point-nonlocal} for scaled effective (a) longitudinal and (b) transverse wave characteristics, $c_e^{L,T}$ and $\gamma_e^{L,T}$, as a function of dimensionless wavenumber $k_{_{L_1}}a$  for the four 3D models of disordered composites of spheres of radius $a$ and $\phi_2=0.25$.
The Poisson ratios of the matrix and dispersed phases are $\nu_1 = 0.4$ and $\nu_2=0.25$, respectively, and the phase contrast ratios are $K_2/K_1 = 10$, $G_2/G_1 = 28$, which correspond to glass beads in an epoxy matrix \cite{sen_bulk_1987}.
Here, $\kLref$ is the longitudinal wavenumber in the reference phase, and $\cLref$ and $\cTref$ are longitudinal and transverse wave speeds, respectively.
The insets in the lower panels are log-log plots of the respective larger panels.
\label{fig:wave-models}}
\end{figure}

Having established the accuracy of the nonlocal strong-contrast approximations, \eqref{eq:Ke_two-point-nonlocal} and \eqref{eq:Ge_two-point-nonlocal}, for simple cubic lattice packings in \sref{sec:vs-approxs}, we now apply them to predict the effective elastodynamic characteristics of the four different disordered models discussed in \sref{sec:models}.
Specifically, we study how the effective elastic moduli [$\fn{K_e}{\kLref}$, $\fn{G_e}{\kLref}$], wave speeds $\fn{c_e^{L,T}}{\kLref}$, and attenuation coefficients $\fn{\gamma_e^{L,T}}{\kLref}$] vary with the microstructure.
For simplicity, we take the matrix phase to be the reference phase (phase 1) and the dispersed phase to be the polarized phase (phase 2).

\Fref{fig:wave-models} shows the scaled effective wave characteristics [i.e., $c_e^L / \cLref$ and $\gamma_e^L /c_e^L$ for longitudinal waves and $c_e^T / \cTref$ and $\gamma_e^T /c_e^T$ for transverse waves] vary with $\kLref$ at fixed phase properties $K_2/K_1=10$, $G_2/G_1=28$, and $\nu_1=0.4$ for the four models.
While all models are effectively lossless (i.e., small values of $\gamma_e$) for a range of wavenumber around the origin, they become increasingly lossy as the wavenumber increases; see the lower panels of \fref{fig:wave-models}.
In the quasistatic regime, as shown in the insets of \fref{fig:wave-models}, hyperuniform and nonhyperuniform exhibit qualitatively different attenuation characteristics [cf. \eqref{eq:atten-characteristics}]: hyperuniform composites generally tend to be less lossy than their nonhyperuniform counterparts.
Remarkably, stealthy hyperuniform media can be perfectly lossless, even well beyond the quasistatic regime; see  \eqref{eq:SHU-transparency}.
Such microstructure-dependence of the effective attenuation behaviors vividly demonstrates that $\gamma_e^{L,T}$ can be engineered by the spatial correlations of composites.

\begin{figure}
\begin{center}
\includegraphics[width=0.6\textwidth]{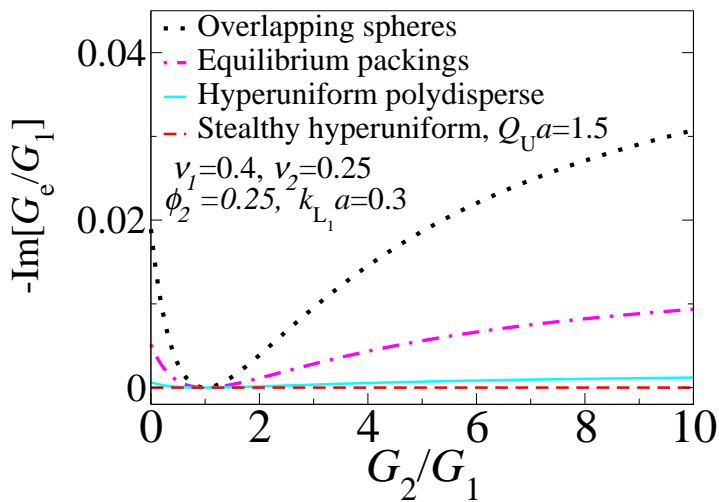}
\end{center}
\caption{Predictions of the nonlocal strong-contrast approximation \eqref{eq:Ge_two-point-nonlocal} for the negative values of the imaginary part of the effective shear modulus $\Imag[\fn{G_e}{\kL}]$ as a function of contrast ratio $G_2/G_1$ of the four disordered models, as per \fref{fig:wave-models}, at volume fraction $\phi_2=0.25$ and wavenumber $\kLref a =0.3$.
The Poisson ratios of the matrix and dispersed phases are fixed at $\nu_1=0.4$ and $\nu_2=0.25$.
\label{fig:Ge-vs-contrast}
}
\end{figure}

We now examine how the imaginary part $\Imag[G_e]$ varies with the contrast ratio $G_2/G_1$ for the disordered models for a given large wavenumber $\kL$ inside the transparency interval (wavenumber ranges where the imaginary parts of the effective bulk and shear moduli vanish) given in \eqref{eq:SHU-transparency} for the stealthy hyperuniform packing.
Here, we fix the phase Poisson ratios to be $\nu_1 = 0.4$ and $\nu_2=0.25$, as we did for the case shown in \fref{fig:wave-models}. 
These results are summarized in \fref{fig:Ge-vs-contrast}.
The disparity in the attenuation characteristics across microstructures widens significantly as the contrast ratio increases.
Clearly, overlapping spheres are the lossiest systems.
Hyperuniform polydisperse packings can be nearly as lossless as stealthy hyperuniform ones.
Unlike the imaginary part, the real part $\Real[G_e]$ is virtually independent of model microstructure and thus is not shown in this work. 
We also do not include the corresponding plot for $K_e$ because its behavior is qualitatively similar to that of $G_e$. 

\begin{figure}[h]
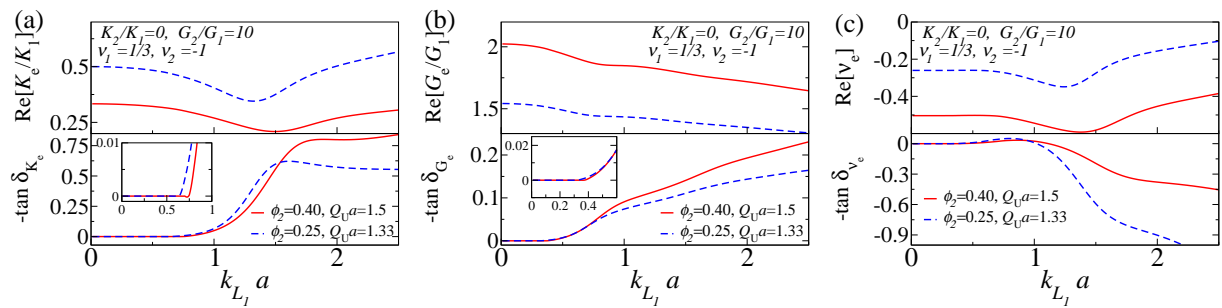

\subfloat{
\includegraphics[width=0.33\textwidth]{Fig7a.eps}
}
\subfloat{
\includegraphics[width=0.33\textwidth]{Fig7b.eps}
}
\subfloat{
\includegraphics[width=0.33\textwidth]{Fig7c.eps}
}
\caption{
Predictions of the nonlocal strong-contrast approximations \eqref{eq:Ke_two-point-nonlocal} and \eqref{eq:Ge_two-point-nonlocal} for the effective (a) bulk $K_e$ and (b) shear $G_e$ moduli, and (c) effective Poisson ratio $\nu_e$ as a function of dimensionless wavenumber $\kLref a$ for 3D stealthy hyperuniform packings of contrast ratio $G_2/G_1 = 10$ at two different packing fractions: $\phi_2=0.4$ and $Q_\mathrm{U} a=1.5$ and $\phi_2=0.25$ and $Q_\mathrm{U} a\approx 1.33$.
The Poisson ratios of the matrix and dispersed phases are $\nu_1=1/3$ and $\nu_2=-1$ (i.e., $K_2/K_1 = 0$), respectively.
In the lower panels of each figure, the negatives values of the corresponding loss tangents [cf. \eqref{eq:loss-factor}] are plotted. 
The insets in (a) and (b) are magnifications of the respective lower panels.
\label{fig:axuetic-cases}
}
\end{figure}

Since stealthy hyperuniform packings exhibit novel physical properties, such as perfect transparency, we further study the effect of packing fraction $\phi_2$ on their effective elastic moduli $\fn{K_e}{\kLref}$ and $\fn{G_e}{\kLref}$ and the effective Poisson ratio $\fn{\nu_e}{\kLref}$. 
Specifically, we are interested in examining stealthy hyperuniform packings consisting of auxetic particles of $\nu_2=-1$ and a matrix phase with $\nu_1=1/3$ and $G_2/G_1=10$. 
Auxetic (negative Poisson ratio) materials laterally dilate (shrink)in response to axial elongation (contraction) \cite{lakes_materials_1993}, and are known to have superior energy-absorbing properties \cite{scarpa2003dynamic}.  
We first generate such packings at a packing fraction $\phi_2=0.4$ and $Q_\mathrm{U}a = 1.5$, as described in \sref{sec:stealthy}.
Without changing particle positions, we then shrink the sphere radii to attain a packing fraction $\phi_2=0.25$, whose stealthy regime is now $Q_\mathrm{U}a \approx 1.33$.

In \fref{fig:axuetic-cases}, we plot the effective bulk and shear moduli as well as the effective Poisson ratio using approximations \eqref{eq:Ke_two-point-nonlocal} and  \eqref{eq:Ge_two-point-nonlocal}, and  \eqref{eq:Poisson ratio}.
To quantifythe damping characteristics of such composites, we also include in this figure  the corresponding {\it loss tangents} defined by
\begin{equation} \label{eq:loss-factor}
\tan \delta_{X_e} \equiv \Imag[X_e]/\Real[X_e],
\end{equation}
for some general effective property $X_e$, which are frequently measured in experiments.
For the bulk and shear moduli, the loss tangents represent the ratios of mechanically attenuated energy to the stored elastic energy \cite{allard_propagation_2009}.
We see that these stealthy dispersions are effectively auxetic, i.e., $\Real[\nu_e]<0$ [see \fref{fig:axuetic-cases}(c)]. 
\Fref{fig:axuetic-cases}(a) reveals that such stealthy auxetic composites have exceptionally large loss tangent values in the intermediate-wavelength regime, compared to typical values ($\lesssim 10^{-1}$ as in the cases in \fref{fig:wave-models}), which implies that they are excellent energy absorbers, as expected. 
The transparency interval (i.e., $\tan \delta_{K_e} = \tan \delta_{G_e} = 0$) is slightly larger for the higher density packing with the higher stealthy cut-off value $Q_U a=1.5a$,
as predicted by \eqref{eq:SHU-transparency}.
The complex Poisson ratio implies that the lateral and axial vibrations are out of phase.
The absolute value of $\tan \delta_{\nu_e}$ is approximately proportional to the difference between the shear and bulk loss factors (i.e., degrees of energy loss due to shear and compression); see Ref. \cite{pritz_poissons_2007}.

\section{Conclusions and Discussion}
\label{sec:discussion}

Closed-form approximations of the effective dynamic elastic moduli derived previously only apply at long wavelengths (quasistatic regime) and for very special  {\it macroscopically isotropic} disordered composite microstructures \cite{kerr_scattering_1992}, namely, nonoverlapping spheres or spheroids in a matrix.
In this work, we have provided the theoretical underpinnings to substantially extend previous work in both its generality and applicability.
First, we derived exact homogenized constitutive relations for the effective dynamic elastic stiffness tensor $\fn{\tens{C}_e}{\kL}$ from first principles that are {\it local in space}. 
Second, our strong-contrast representation of $\fn{\tens{C}_e}{\kL}$ exactly accounts for complete microstructural information ($n$-point correlation functions for $n\geq 1$) for {\it general microstructures} and hence multiple scattering to all orders in the quasistatic regime.
Third, we extracted from the exact expansions accurate local closed-form approximate formulas for $\fn{K_e}{\kL}$ and $\fn{G_e}{\kL}$, relations  \eqref{eq:Ke_two-point-nonlocal} and \eqref{eq:Ge_two-point-nonlocal}, which are resummed representations of the exact expansions  that incorporate microstructural information through the spectral density $\spD{\vect{Q}}$, which is easily ascertained for general microstructures either theoretically, computationally or via scattering experiments.
Depending on whether the high-stiffness phase percolates or not, the wide class of microstructures that we can treat includes particulate media consisting of identical or polydisperse particles of general shape (ellipsoids, cubes, cylinders, polyhedra)
with prescribed orientations that may or not {\it overlap}, cellular networks as well as media without well-defined inclusions (\sref{sec:local-str-cont}).
Fourth, we extended these local approximations beyond the quasistatic regime by postulating nonlocal formulas based on the similarities between electrodynamic and elastodynamic problems and our rigorous formulation of the nonlocal effective dynamic dielectric properties \cite{torquato_nonlocal_2020}.
We carried out precise full-waveform elastodynamic simulations for certain 3D benchmark models to validate the accuracy of our nonlocal formulas for wavenumbers well beyond the quasistatic regime, i.e.,  $0 \le  \kL \ell \lesssim 1$
(where $\ell$ is a characteristic heterogeneity length scale).

Having verified the accuracy of the postulated strong-contrast approximations \eqref{eq:Ke_two-point-nonlocal} and \eqref{eq:Ge_two-point-nonlocal} for dispersions, we then applied them to the four disordered model microstructures in three dimensions (both nonhyperuniform and hyperuniform) to investigate the microstructure-dependence of the effective elastic wave characteristics.
	We demonstrated that disordered hyperuniform media are generally less lossy than their nonhyperuniform counterparts.
	We also found that our approximations predict that disordered hyperuniform media possess a transparency wavenumber interval \eqref{eq:SHU-transparency} around which most nonhyperuniform media exhibit strong attenuation.
	We note that using finite-element method calculations and supercell techniques, Gkantzounis, Amoah, and Florescu \cite{gkantzounis_hyperuniform_2017} showed that 2D stealthy hyperuniform packings should exhibit a transparency interval for elastic waves, which are qualitatively consistent with our predictions.

The accuracy of our nonlocal closed-form formulas has important practical implications since one can now use them to accurately and efficiently predict the effective wave characteristics well beyond the quasistatic regime of a wide class of composite microstructures without carrying out computationally expensive full-blown simulations. 
Thus, our nonlocal formulas can be used to accelerate the discovery of novel elastodynamic composites by appropriate tailoring of the spectral densities and then constructing the corresponding microstructures by using Fourier-space inverse methods \cite{Chen2017}.
For example, from our findings, it is clear that stealthy disordered particulate media can be utilized as low-pass filters that transmit elastic waves ``isotropically" up to a selected wavenumber.
Of course, one could also explore the design space of effective elastic wave properties of nonhyperuniform
disordered composite media for potential applications.

There are interesting open problems for future exploration.
Could the exact local strong-contrast expansions, such as \eqref{eq:exp-anisotropic} and \eqref{eq:exp-isotropic}, be generalized to the cases in which the mass densities of both phases are different, i.e., $\rho_1\neq \rho_2$?
This is a highly nontrivial extension. 
One possible approach to answer this question is to introduce the concept of the dynamic matrix, which is used to derive dispersion relations for elastic waves in simple harmonic lattices \cite{Ashcroft} in order to separate the local mass density $\fn{\rho}{\vect{x}}$ and displacement field $\fn{\vect{u}}{\vect{x}}$.
Another challenging problem is the derivation of strong-contrast expansions of the effective dynamic elastic moduli that are nonlocal in space from the first principles in the manner obtained for the electromagnetic problem \cite{torquato_nonlocal_2020}.
This problem is also quite challenging partly because, unlike the electromagnetic waves, one needs to account for the interplay between longitudinal and transverse propagation modes of elastic waves at a given frequency.
Finally, it desirable to formulate full-waveform elastodynamic  simulations for two-phase media that are more efficient than the dynamic FFT scheme used here. 

\section*{Acknowledgements}
The authors gratefully acknowledge the
support of the Air Force Office of Scientific Research Program on
Mechanics of Multifunctional Materials and Microsystems under
award No. FA9550-18-1-0514.

{\appendix

\section{Strong-Contrast Approximations at the Three-Point Level}
\label{app:3pt}

Here, we explicitly present strong-contrast approximations at the three-point level for the effective dynamic elastic moduli in the quasistatic regime.
Specifically, truncating \eqref{eq:exp-bulk} and \eqref{eq:exp-shear} at the three-point level and solving the left-hand sides of these truncated series for $K_e$ and $G_e$, respectively, yields 
\begin{eqnarray}
\quad\fl \frac{\fn{K_e}{\kL}}{K_q} =& 1+ \left[\frac{\cL^2}{\cL^2 - 2(1-1/d)\cT^2}\right]
\frac{{\phi_p}^2 \kappa_{pq}}{\phi_p (1-\phi_p \kappa_{pq}) - [\fn{C_2^{(p)}}{\kL}+\fn{C_3 ^{(p)}}{\kL} ]}, \label{eq:Ke_three-point}  \\
\quad\fl\frac{\fn{G_e}{\kL}}{G_q} =& 1+ \left[\frac{d(d+2)\cL^2 /2}{d\cL^2 +2 \cT^2}\right]
\frac{{\phi_p}^2 \mu_{pq}}{\phi_p (1-\phi_p \mu_{pq})- [\fn{D_2^{(p)}}{\kL} +\fn{D_3^{(p)}}{\kL}]},\label{eq:Ge_three-point} 
\end{eqnarray}
where the explicit formulas for the three-point parameters $\fn{C_3^{(p)}}{\kL}$ and $\fn{D_3^{(p)}}{\kL}$ are given respectively as
	\begin{eqnarray}
\fl \fn{C_3 ^{(p)}}{\kL} =& -\frac{\kL^{d+2}}{2^{2+d}\pi^{d-2}}\int\hspace{-8pt}\int \frac{\dd{\vect{r}}}{r^{d/2-1}}\frac{\dd{\vect{s}}}{s^{d/2-1}} 
\Bigg\{
\kappa_{pq}\mu_{pq} \frac{(d+2)\cT^2}{d \cL^2 + 2\cT^2} \Hankel{d/2+1}{\kL r} \Hankel{d/2+1}{\kL s}\fn{\hat{P}_2}{\uvect{r}\cdot\uvect{s}}
	\nonumber \\
\fl&+ {\kappa_{pq}}^2 \Hankel{d/2-1}{\kL r} \Hankel{d/2-1}{\kL s} \Bigg\} \left[\fn{S_3 ^{(p)}}{\vect{r},\vect{s},\vect{t}} - \frac{\fn{S_2 ^{(p)}}{\vect{r}}\fn{S_2 ^{(p)}}{\vect{s}}}{\phi_p}\right], \label{eq:C3-explicit}	\\
\fl	\fn{D_3 ^{(p)}}{\kL} =& \frac{2}{d-1} \frac{\kL^{d+2}}{2^{2+d}\pi^{d-2}}\frac{\cT^2 \mu_{pq}}{d\cL^2 + 2\cT^2}\int\hspace{-8pt}\int \frac{\dd{\vect{r}}}{r^{d/2-1}}\frac{\dd{\vect{s}}}{s^{d/2-1}} \Bigg(-\kappa_{pq} \fn{\hat{P}_2}{\uvect{r}\cdot\vect{s}} \Hankel{d/2+1}{\kL r} \Hankel{d/2+1}{\kL s}  \nonumber \\
\fl&- \frac{(d+2)\cT^2 \mu_{pq}}{d\cL^2 + 2\cT^2}
\Bigg\{
d^2\fn{\hat{P}_4}{\uvect{r}\cdot\vect{s}}
\left[\Hankel{d/2+3}{\kL r} - \frac{\Hankel{d/2+3}{\kT r}}{\left(\cT/\cL\right)^{d/2+3}}\right] 
\nonumber \\
&\quad\times
\left[\Hankel{d/2+3}{\kL s} - \frac{\Hankel{d/2+3}{\kT s}}{\left(\cT/\cL\right)^{d/2+3}}\right] 
\nonumber \\
\fl&+ 
\frac{d-2}{4(d+4)}\fn{\hat{P}_2}{\uvect{r}\cdot\vect{s}} 
\left[4\Hankel{d/2+1}{\kL r} + d\frac{\Hankel{d/2+1}{\kT r}}{\left(\cT/\cL\right)^{d/2+3}}\right]
\nonumber \\
&\quad\times 
\left[4\Hankel{d/2+1}{\kL s} + d \frac{\Hankel{d/2+1}{\kT s}}{\left(\cT/\cL\right)^{d/2+3}}\right] 
\nonumber \\
\fl&+
\frac{d-1}{2(d+2)}
\left[2\Hankel{d/2-1}{\kL r} +d \frac{\Hankel{d/2-1}{\kT r}}{\left(\cT/\cL\right)^{d/2+3}}\right] 
\left[2\Hankel{d/2-1}{\kL s} + d\frac{\Hankel{d/2-1}{\kT s}}{\left(\cT/\cL\right)^{d/2+3}}\right]
\Bigg\}\Bigg)
\nonumber \\
\fl&\times \left[\fn{S_3 ^{(p)}}{\vect{r},\vect{s},\vect{t}} - \frac{\fn{S_2 ^{(p)}}{\vect{r}}\fn{S_2 ^{(p)}}{\vect{s}}}{\phi_p}\right].	\label{eq:D3-explicit}
\end{eqnarray}
Here $\vect{t}\equiv \vect{r}-\vect{s}$, and 
	\begin{eqnarray}
	\fn{\hat{P}_4}{t} &\equiv t^4-\frac{6}{d+4}t^2+\frac{3}{(d+2)(d+4)} \label{eq:Legendre_4th},\\
	\fn{\hat{P}_2}{t} &\equiv dt^2-1. \label{eq:Legendre_2nd}
	\end{eqnarray}	
In the static limit ($\omega = 0$), the three-point parameters \eqref{eq:C3-explicit} and \eqref{eq:D3-explicit} reduce to
\begin{eqnarray}
\fl\fn{C_3^{(p)}}{0} & = \frac{(d+2)\cT^2}{d \cL^2 + 2\cT^2} \kappa_{pq}\mu_{pq} \frac{\fn{\Gamma}{d/2+1}^2}{\pi^d} \int\hspace{-8pt}\int \frac{\dd{\vect{r}}}{r^d}\frac{\dd{\vect{s}}}{s^d} 
\fn{\hat{P}_2}{\uvect{r}\cdot\uvect{s}}
 \left[\fn{S_3 ^{(p)}}{\vect{r},\vect{s}} - \frac{\fn{S_2 ^{(p)}}{\vect{r}}\fn{S_2 ^{(p)}}{\vect{s}}}{\phi_p}\right] 
 \nonumber \\
\fl & = \frac{(d-1)(d+2)\cT^2}{d \cL^2 + 2\cT^2} \kappa_{pq}\mu_{pq}  \phi_q \phi_p \zeta_p,
 \label{eq:C3-static}\\
\fl\fn{D_3 ^{(p)}}{0} &= \frac{2}{d-1}\frac{\cT^2}{d \cL^2 + 2\cT^2} \mu_{pq} \frac{\fn{\Gamma}{d/2+1}^2}{\pi^d}
\int\hspace{-8pt}\int \frac{\dd{\vect{r}}}{r^d}\frac{\dd{\vect{s}}}{s^d} 
\Bigg(
	\kappa_{pq}\fn{\hat{P}_2}{\uvect{r}\cdot\uvect{s}} 
	+ \frac{(d+2)\cT^2}{d \cL^2 + 2\cT^2} \mu_{pq} 
	\nonumber \\
\fl	&\quad \times
		\Bigg\{
		 \frac{d-2}{d+4} \frac{(d\cL^2 +4\cT^2)^2}{4{\cT}^4} \fn{\hat{P}_2}{\uvect{r}\cdot\uvect{s}}  
		+\frac{d^2(d+2)^2}{4}\frac{(\cL^2-\cT^2)^2}{{\cT}^4} \fn{\hat{P}_4}{\uvect{r}\cdot\uvect{s}}		
		\Bigg\} 
\Bigg) \nonumber \\
\fl &\quad \times
	\left[\fn{S_3 ^{(p)}}{\vect{r},\vect{s},\vect{t}} - \frac{\fn{S_2 ^{(p)}}{\vect{r}}\fn{S_2 ^{(p)}}{\vect{s}}}{\phi_p}\right] \nonumber \\
  &= \frac{2\cT^2 \phi_p\phi_q \mu_{pq}}{d \cL^2 + 2\cT^2}  \Bigg\{\kappa_{pq} \zeta_p
+ \frac{(d+2)\cT^2 \mu_{pq}}{d \cL^2 + 2\cT^2}  \Bigg[\frac{d(d-2)}{4} \frac{2\cL^2 - (1-4/d)\cT^2}{\cT^2} \zeta_p 
\nonumber \\
 &\quad + \frac{d^3}{4(d+2)} \frac{(\cL^2-\cT^2)^2}{\cT^4} \eta_p \Bigg]
\Bigg\}	,\label{eq:D3-static}
\end{eqnarray}
where the parameters $\eta_p$ and $\zeta_p$ lie in a closed interval $[0,1]$; see Ref. \cite{Torquato_RHM} and reference therein.

\section{Gaunaurd-\"{U}berall Approximation}
\label{sec:approxs}

Here we state explicit formulas for the Gaunaurd-\"{U}berall approximation (GUA) for the effective dynamic elastic moduli of an isotropic medium composed of identical spheres of radius $a$ in the quasistatic regime \cite{gaunaurd_resonance_1983, kerr_scattering_1992}. 
The particles are in phase 2 of mass density $\rho_2$ and elastic moduli $K_2$ and $G_2$, and they are embedded in a matrix phase of $\rho_1$, $K_1$ and $G_1$.

Since the GUA accounts solely for scatterings from a single particle in the mean-field treatments, it can be regarded as the elastodynamic counterpart of the Maxwell-Garnett approximation \cite{ruppin_evaluation_2000}.
The explicit formulas for the effective bulk and shear moduli are given respectively as \cite{kerr_scattering_1992} 
\begin{eqnarray}
&\frac{\fn{K_e}{\kLref} -K_1}{\fn{K_e}{\kLref} + 4G_1 /3 - [\Gamma_e R^2 (\kLref a)^2 - i R^3 (\kLref a)^3 (\fn{K_e}{\kLref} - K_1)]/3} 
\nonumber \\
& \quad\quad\quad =  \frac{\phi_2 \kappa_{21}}{1 - [\Gamma_2 (\kLref a)^2/(3K_2 + 4G_1) - i(\kLref a)^3 \kappa_{21} /3]}, \label{eq:GU_model_Ke}
\end{eqnarray}
\begin{eqnarray}
&\frac{\fn{G_e}{\kLref} -G_1}{\fn{G_e}{\kLref} + \frac{[3 K_1 /2 +4G_1 /3]G_1}{K_1 + 2G_1}} = \phi_2 \mu_{21} \label{eq:GU_model_Ge},
\end{eqnarray}
where $\kappa_{21}$ and $\mu_{21}$ are given in \eqref{eq:bulk_modulus_polarizability} and \eqref{eq:shear_modulus_polarizability}, respectively, $\rho_e = \rho_1 +\phi_2 (\rho_2 - \rho_1)$, $R$ represents the radius of a specimen, which is often set to be zero \cite{kerr_scattering_1992}, and $\Gamma_i$ (for $i=2, e$) are given as
\begin{eqnarray}
\Gamma_i & = K_1 -\frac{3}{2}K_i - \frac{2}{3}G_1  + \frac{\rho_i}{2 \rho_1} \frac{3K_1 + 4G_1}{3K_i + 4G_i} \left[ K_i +\frac{4}{5}(G_1+\frac{2}{3}G_i)\right].
\end{eqnarray}

\section{Properties of the Attenuation Functions}
\label{sec:prop-atten}

Here, we present asymptotic behaviors of both attenuation functions $\fn{\mathcal{F}}{Q}$ and $\fn{F}{Q}$, defined by \eqref{eq:local-F-real} and \eqref{eq:nonlocal-F}, respectively, for a statistically isotropic composite.
We then briefly discuss the transparency condition \eqref{eq:SHU-transparency} for stealthy hyperuniform media.
Both are functionals of the spectral density $\spD{Q}$ and identical in the quasistatic regime.
Specifically, assuming that the spectral density has the power-law scaling $\spD{Q}\sim Q^\alpha$ as $Q\to 0$, the attenuation functions become 
\begin{eqnarray}
\quad\fl\Imag[\fn{\mathcal{F}}{Q}] = \Imag[\fn{F}{Q}] &\sim 
\begin{cond}
Q^d,			&\mathrm{nonhyperuniform }(\alpha=0)\\
Q^{d+\alpha}, 	&\mathrm{hyperuniform }(\alpha>0)
\end{cond},&\mathrm{as~}Q\to 0, \label{eq:small-Q-Im_atten_fn}
\\
\quad\fl\Real[\fn{\mathcal{F}}{Q}]=\Real[\fn{F}{Q}] &\sim Q^2,& \mathrm{ as~}Q\to 0,
\label{eq:small-Q-Re_atten_fn}
\end{eqnarray}
where $\alpha=0$ for nonhyperuniform systems, and $\alpha>0$ for hyperuniform systems.
In the large-$Q$ regime, both types of attenuation functions exhibit considerably different scalings:
\begin{eqnarray}
\quad\fl&\Imag[\fn{\mathcal{F}}{Q}]\sim Q^{-1},
&\Real[\fn{\mathcal{F}}{Q}]\to \frac{2^{d/2}\fn{\Gamma}{d/2}}{\pi}\phi_p
(1-\phi_p) ~~(>0),
~~\mathrm{as~}Q\to\infty \label{eq:large-Q-local-atten-fn}\\
\quad\fl&\Imag[\fn{F}{Q}]\sim Q,
&\Real[\fn{F}{Q}]\to \mathrm{const.} ~~(<0),
\mathrm{as~}Q\to\infty, \label{eq:large-Q-nonlocal-atten-fn}
\end{eqnarray}
regardless of whether the composites are hyperuniform or not.
The reader is referred to Ref. \cite{torquato_nonlocal_2020} for derivations.

A two-phase composite is effectively lossless for elastic waves at a given frequency $\omega$ if and only if $\Imag[\fn{K_e}{\kL}]=0$ and $\Imag[\fn{G_e}{\kL}]=0$, which are equivalent to $\Imag[\fn{F}{\kL}]=\Imag[\fn{F}{\kT}]=0$ when using the nonlocal approximations \eqref{eq:Ke_two-point-nonlocal} and \eqref{eq:Ge_two-point-nonlocal}.
One can show that these conditions are satisfied in the transparency interval \eqref{eq:SHU-transparency} for stealthy hyperuniform media [cf. \eqref{eq:SHU-condition}].
}

\section*{References}
\providecommand{\newblock}{}

\end{document}